\documentclass[useAMS,usenatbib]{mn2e}
\usepackage{graphicx}
\usepackage{amssymb}
\usepackage{amsmath,tabu,physics}
\usepackage{subfigure}
\usepackage{ifpdf}
\usepackage{color}
\usepackage{gensymb,comment, tabularx}
\usepackage{float}

\title[Multi-tracer dynamical models of WLM]{Joint Gas and Stellar Dynamical Models of WLM: An isolated dwarf galaxy within a cored, prolate DM halo}
\author[Leung et al.]
{Gigi Y.\,C. Leung$^{1}$\thanks{Email: leung@mpia.de}, Ryan Leaman$^{1}$, Giuseppina Battaglia$^{2,3}$, Glenn van de Ven$^{4}$, 
\newauthor Alyson M. Brooks$^{5}$, Jorge Pe\~{n}arrubia$^{6}$ and Kim A. Venn$^{7}$\\
$^{1}$Max-Planck Institut f\"ur Astronomie, K\"onigstuhl 17, D-69117 Heidelberg, Germany\\
$^{2}$Instituto de Astrof\'{i}sica de Canarias, Calle V\'{i}a L\'{a}ctea s/n, E-38205 La Laguna, Tenerife, Spain\\
$^{3}$Departamento de Astrof\'{i}sica, Universidad de La Laguna, E-38200 La Laguna, Tenerife, Spain\\
$^{4}$Department of Astrophysics, University of Vienna, T\"urkenschanzstrasse 17, 1180 Vienna, Austria\\
$^{5}$Department of Physics \& Astronomy, Rutgers, The State University of New Jersey, 136 Frelinghuysen Rd, Piscataway, NJ 08854, USA\\
$^{6}$Institute for Astronomy, University of Edinburgh, Royal Observatory, Blackford Hill, Edinburgh EH9 3HJ\\
$^{7}$Department of Physics and Astronomy, University of Victoria, Victoria, BC V8W 3P2, Canada
}
\date{Accepted 2019. Received 2019; in original form 2019}

\pagerange{\pageref{firstpage}--\pageref{lastpage}} \pubyear{2019}

\begin{document}

\label{firstpage}

\maketitle

\begin{abstract}
We present multi-tracer dynamical models of the low mass ($M_{*} \sim 10^{7}$), isolated dwarf irregular galaxy WLM in order to simultaneously constrain the inner slope of the dark matter (DM) halo density profile ($\gamma$) and flattening ($q_\mathrm{DM}$), and the stellar orbital anisotropy ($\beta_{z}, \beta_{r}$).  For the first time, we show how jointly constraining the mass distribution from the HI gas rotation curve and solving the Jeans' equations with discrete stellar kinematics leads to a factor of $\sim2$ reduction in the uncertainties on  $\gamma$.  The mass-anisotropy degeneracy is also partially broken, leading to reductions on uncertainty by $\sim 30\%$ on $M_\mathrm{vir}$ (and $\sim 70\%$ at the half-light radius) and $\sim 25\%$ on anisotropy. Our inferred value of $\gamma = 0.3 \pm 0.1$ is robust to the halo geometry, and in excellent agreement with predictions of stellar feedback driven DM core creation. The derived prolate geometry of the DM halo with $q_\mathrm{DM} = 2 \pm 1$ is consistent with $\Lambda$CDM simulations of dwarf galaxy halos. While self-interacting DM (SIDM) models with $\sigma/m_{X} \sim 0.6$ can reproduce this cored DM profile, the interaction events may sphericalise the halo. The simultaneously cored and prolate DM halo may therefore present a challenge for SIDM. Finally we find that the radial profile of stellar anisotropy in WLM ($\beta_{r}$) follows a nearly identical trend of increasing tangential anisotropy to the classical dSphs, Fornax and Sculptor. Given WLM's orbital history, this result may call into question whether such anisotropy is a consequence of tidal stripping in only one pericentric passage or if it instead is a feature of the largely self-similar formation and evolutionary pathways for some dwarf galaxies.
\end{abstract}


\section{Introduction}

The shape and radial density profile of dark matter (DM) halos provides a window into the nature of dark matter, and the efficiency of baryonic feedback processes which influence the galaxies residing in these halos \citep[e.g.,][]{dic14}. For instance, dark-matter only cosmological and $N$-body simulations have shown that, under the $\Lambda$CDM cosmological framework, the dark matter haloes around galaxies follow a cuspy density profile characterised by an NFW profile \citep[e.g.,][]{nfw96, dut14}.  Hydrodynamic simulations which incorporate baryonic feedback from stars and AGN find that not only are the star formation properties altered, but also the repeated ejection of gas from central regions of low mass galaxies can result in a reduction of the inner cumulative baryonic and dark mass distribution; e.g., \citealt{nav96,read05,Maschenko06,pen12,pon12}).

The DM halo properties may hence be correlated with the baryonic content of the galaxies. For example, \citet{dic14} show that the inner slope of the dark matter haloes correlates with the stellar-mass-to-halo-mass ratio in their simulated galaxies, and \citet{read16} showed with hydrodynamical simulations that the core size of the dark matter haloes in dwarf galaxies generally correlates with the half-light radii of the stellar component.  Significant variation in the predicted range of dark matter fractions (in terms of mass with respect to the total mass of the galaxy) is seen either directly from cosmological zoom-in simulations \citep{bro15}, or from abundance matching predictions \citep[e.g.,][]{leau12,saw13}.  Understanding this stochasticity is therefore crucial to gain a better understanding of the efficiency with which baryonic feedback can suppress star formation - and simultaneously alter the initial dark matter halo profiles.

Being the most dark-matter dominated objects in the universe, dwarf galaxies act as prime laboratories for testing the impact of baryonic feedback and the nature of dark matter. Various techniques have been adopted to infer the relative contribution of stellar and dark components in low mass galaxies. For example, decomposition of rotation curves obtained from HI kinematics has been used to study the fractional amount of dark matter in low mass galaxies \citep[e.g.][]{lel10, swa11, adams14, katz17}. These results typically found that despite the uncertainties in stellar mass-to-light ratios, the baryonic mass was a small fraction of that necessary to reproduce the circular velocity profiles. These objects thus can provide a stringent test also on the nature of dark matter and/or non-Newtonian dynamics \citep[e.g.][]{lel10, mcg13, vog14}.

Many of these same studies found that that the rotation curves of nearby dwarf galaxies have inner density or cumulative mass profiles that are shallower/smaller than the cosmologically predicted cuspy NFW profile of DM-only simulations \citep[e.g.][]{oh11, adams14, bro15}. This instead may be a signature of the aforementioned stellar feedback driven DM coring, which is likely to be most effective in low mass galaxies with shallow potentials, but still significant star formation \citep{read05, dic14}. A challenge in assessing this scenario is that the asymptotic slope of the DM density profile is difficult to infer and depends on the particular mass profile assumed.  Several studies instead have characterized the density profile slope at a fixed physical scale, or fraction of the virial radius or stellar half-light radius \citep[e.g.][]{Hague13,Read19,Li19}. The need for isolated galaxies which have not been environmentally stripped of gas, and of a high enough mass such that the HI rotation is measurable, means that these techniques have not typically been used for the most nearby Local Group or low mass satellite galaxies of the MW. 

In those systems, stellar kinematics are predominantly used to measure the dark matter density profiles (either estimating an asymptotic slope, or the slope of the profile at fixed radius), through the virial mass estimates \citep{walk11}, the Jeans equations \citep[e.g.][]{lok09, zhu16} or Schwarzschild models \citep[e.g.][]{bre13, kow18}. Measuring the mass profile from stellar kinematics suffers from uncertainties associated with the unknown velocity anisotropy, known as the mass-anisotropy degeneracy. To break the mass-anisotropy degeneracy, the higher order moment (kurtosis) has been have utilised \citep[e.g.][]{lok09,bre13b}.  As this degeneracy is found to have a spatial dependence and is minimal at the half-light radius \citep[e.g.][]{wolf2010, err18}, other authors have separated stellar kinematics into populations of different chemistry with different spatial and kinematical distributions to serve as a lever arm to understand the host potential \citep{bat08, walk11, zhu16}. The constraints on the inner slopes of the dark matter density profiles by stellar kinematics alone is difficult however. For example, while \citet{walk11} can exclude a completely cuspy NFW profile in the dark matter halo of Fornax with up to 96\% significance, the estimated inner slope of $\gamma=0.5\pm0.4$ (where $\gamma$ parametrises the inner slope of a generalised NFW profile, with $\gamma=0$ corresponding to a cored profile and $\gamma=1$ an NFW profile) still has a large uncertainty. Similarly, even with a discrete Jeans model on two chemically distinct population, \citet{zhu16} could only constrain the inner slope of the dark matter halo of Sculptor to be within $\gamma=0.5\pm0.3$ . In another study \citet{kow18} showed that while a cored profile is preferred by their models for Fornax, cuspy NFW and Einasto profiles fall within the 1$\sigma$ uncertainties. Using orbit-based dynamical models extending also to higher moments, \cite{bre13b} found that the stellar kinematics of four dSphs Fornax, Sculptor, Carina and Sextans are compatible with both cuspy and cored DM haloes. Given the difficulties in robustly inferring the profile shape through single or even multiple population stellar tracers, It is desirable to study low mass dwarf galaxies with multiple kinematic tracers (e.g., gas and stars) with new analysis methods.

Combining a collisional gas tracer with discrete kinematic stellar tracers in principle should offer a more robust characterisation of the host potential.  Despite their different orbital structure, the gas and the stellar kinematics should consistently trace the same potential when all sources of orbital energy are accounted for.  Combining observations of stars and gas kinematics in the same galaxy then offers a way to break the mass anisotropy degeneracy and better characterise the dark matter halo properties. Also, while stellar kinematics typically only allows for fitting the mass enclosed within the stellar radii of the chemodynamical components, cold gas provides kinematic information over a much larger radial range. Observations of gas and stars in homogenous observations of a variety of galaxies were presented in \cite{leung18} and for $8.5<\log L_\star<9.5$ dwarf galaxies in \cite{adams14}. However neither of these studies leveraged the tracers simultaneously to measure halo properties from the combined information of both tracers. Nevertheless there appears great promise in exploiting the simultaneous tracers for galaxies where both exist.

Apart from constraints on the underlying gravitational potential, proper modelling of the stellar kinematics can recover their orbit distribution in the galaxy. The shape of the velocity ellipsoid, often parameterised in terms of an anisotropy parameter such as $\beta_{\phi} = 1 - (\sigma_{\phi}/\sigma_{R})^{2}$ (where $\sigma_\phi$ and $\sigma_R$ are the velocity dispersions along the azimuthal and radial direction respectively in a cylindrical coordinates) provides an understanding of the relative amount of random motions in the tangential and radial directions. These quantities may be intimately tied to the formation and evolutionary pathways of the dwarf galaxies - either environmental or secular.  Characterising the anisotropy profiles of dwarfs in the Local group is particularly helpful in understanding any evolutionary connection between dwarf irregulars (dIrr) and dwarf spheroidals (dSph).

For example, predictions of simple dissipationless collapse result in an isotropic core surrounded by an envelope of more radial orbits \citep{valb82}. While dwarfs with sufficiently radially anisotropic orbits may have undergone bar formation, which after subsequent buckling and excitation of bending modes, can result in significant morphological transformations \citep[e.g.,][]{may06,raha91}. Tidally stripped galaxies are thought to have strongly tangential anisotropy in their outer regions as the radial orbits with larger apocentres may be preferentially removed \citep{kli09}.

The tangential velocity anisotropies found in dSphs \citep[e.g.,][]{zhu16, kow18} may agree with some tidal transformation scenarios \citep[e.g.,][]{kli09}, where dIrrs are tidally disturbed and lose their gas and form dSphs, leaving behind a tangential stellar anisotropy distribution for the resultant dSph. This scenario however may be challenged by the existence of transition type dwarfs in isolation \citep[e.g. VV124,][]{vor59}, and the suggested similarities in the  ratio of ordered to random motion $V/\sigma$ between dSphs and dIrrs \citep{whee17}. Comparable estimates of velocity anisotropy in isolated dwarf galaxies, yet to be determined, would serve as a crucial control sample, and help differentiate if this signature is caused by environmental effects, or rather something intrinsic to the formation of any low mass dwarf. 

The recovery of the stellar anisotropy is not trivial and several degeneracies work to prevent accurate understanding of the stellar orbital, or dark matter halo properties. In addition to the aforementioned mass-anisotropy degeneracy \citep[e.g.,][]{bin87} have shown that the derived anisotropy is highly degenerate with the DM halo geometry. This then means that another parameter, the halo flattening $q_\mathrm{DM}$, needs to be introduced in dynamical models in order to recover an unbiased estimate of $\beta$. In a handful of MW mass galaxies,  inference of the DM halo flattening has been produced from HI gas kinematics and structure \citep[e.g.][]{Obrien10,Khoperskov14,Peters17}, with the results depending on the viewing angle and configuration (edge-on, polar ring) as well as assumptions on the gas opacity.   For low mass nearby galaxies, while attempts in dynamical modelling incorporating a halo flattening with fixed anisotropy has been attempted \citep[e.g.][]{hay12}, incorporating variable DM profiles ($\gamma, q_{DM}$) and anisotropy simultaneously, has not been done as the constraints on parameters of interest get understandably poorer with the increasing (but necessary) model complexity.  The necessity of understanding DM in low mass dwarfs, breaking anisotropy and halo property degeneracies, and testing the intrinsic orbit structure of \textit{isolated} dwarf galaxies, clearly motivates the need for a new analysis techniques and observations.

In this work we demonstrate a promising way forward, by jointly modelling the stellar and gaseous kinematics in dwarf galaxies which have both resolved stellar kinematics, and well behaved HI gas rotation curves. With an alternate constraint on the galaxy's potential from the gas rotation curve, the stellar anisotropy estimate should be improved.  A second necessary aspect of the modelling is to flexibly parameterise the DM halo's shape and inner density profile slope.

Often, the nature of the dwarf galaxies prevents observable stellar and gaseous tracers from co-existing, such as in the case of the nearby, quenched dSphs, or the low gas fraction transition dwarfs, or the observational cost of getting stellar kinematics in gas rich distant dwarf irregulars. In the few dwarfs where both resolved stellar and gaseous kinematics have be observed \citep{lea12, kir14}, the dynamical mass estimates from both tracers individually show agreement - provided contributions of non-circular motions are taken into account (e.g., \citealt{hinz01, read17}), a joint dynamical model leveraging both tracers simultaneously has however not yet been attempted.

One of the prime targets, which is near enough for obtaining sufficient stellar kinematics, and massive enough to have a well defined gaseous rotation curve, is the isolated dIrr Wolf-Lundmark-Melotte (WLM; \citet{wolf10, mel26}) . WLM lies at a distance of $\sim1$\,Mpc from both the Milky Way and M31. The distance between WLM and its nearest neighbour, a low-mass dSph Cetus (enclosed mass at half-light radius of $M\sim4\times10^7M_\odot$; \citet{kir14}), is $\sim$250\,kpc \citep{whit99}. With a velocity of $v_\mathrm{LG}\sim-32\,$km\,s$^{-1}$ towards the barycentre of the Local Group, \citet{lea12}  suggested that WLM has just passed its apocentre and would have at most one pericentre passage in its lifetime, which occurred at least 11\,Gyrs ago. Constructing our proposed dynamical model of a dwarf galaxy in such extreme isolation would provide a null test on the effects of external influences, such as tides and ram pressure, and provide one of the most detailed views of the DM halo and orbit structure of a low mass dwarf. Also, WLM's isolated location (together with its comprehensive constraints on thickness, stellar dispersion and circular velocity) renders it as an excellent test case for modified gravity, as external field effects cannot be invoked.

In the following, we first describe our HI and stellar data in Section \ref{sect_data}. We then lay out the observational and model ingredients, including our construction of the dynamical model, the spatial distribution of the kinematic tracers, the baryonic and dark matter density profile, and the steps of our parameter estimation in Section \ref{sect_model}. We present the obtained dark matter halo parameters and velocity anisotropies of WLM in Section \ref{sect_res}. In Section \ref{sect_diss}, we discuss the cosmological implications of the derived dark matter halo profile and flattening, as well as the meaning of the derived orbital structure in terms of the evolution of dwarf galaxies. We conclude in Section \ref{sect_con}.

\section{Data}\label{sect_data}
\subsection{HI interferometric data}\label{sect_HIdata}
We have taken the HI integrated intensity map and the circular velocity $V_\mathrm{c}$ estimated using HI kinematics originally presented in \citet{kep07} and re-analysed by \citet{ior17}. The interferometric data is taken using the Very Large Array, with a beam size of $\sim10\arcsec$ and a velocity resolution of $\sim2.6\,$km\,s$^{-1}$. The integrated intensity map is shown as black contours on the left panel of Figure \ref{fig_HIstarobs}, the velocity map from which the circular velocities are derived from is shown on the right panel of Figure \ref{fig_HIstarobs}. From the velocity map, \citet{ior17} have derived an inclination of 74$\degree$ and a position angle of 174$\degree$, which we adopt throughout the whole paper. Their derived $V_\mathrm{c}$ is shown in Figure \ref{fig_HIvc}.

\subsection{Photometric Data}
The I band photometry was obtained using the INT Wide Field Camera and presented initially in \citet{mcc05} and covers a $36^{'}\times 36^{'}$ field of view.  We used the resolved radial stellar number density profiles constructed from this data and presented in \cite{lea12} in both I band, and the JHK photometric observations of \cite{tat11}.  We refer the reader to \cite{lea12} for details of the profile construction.

In addition we utilise photometric observations in the I band taken with the MOSAIC-II imager formerly installed on the 4m Blanco telescope at CTIO.  These observations were taken in excellent seeing conditions ($\sim 0.8^{"}$) on September 11 -12, 2009 (PI: Leaman 2009B-0337).  The CCD has a pixel scale of $0.27^{"}$/pixel and the images were processed and coadded through the NOAO Science Archive pipelines.  The co-added stacked image which was used to build the stellar contribution to the mass distribution, covers a field of view of  $0.63\times0.67$ degrees.  Further details of the observations and reductions will be presented in Hughes et al. (in prep.).

\subsection{Resolved stellar spectroscopy}
\begin{figure}
\begin{center}
\includegraphics[width=0.525\textwidth, trim = 20 0 400 20]{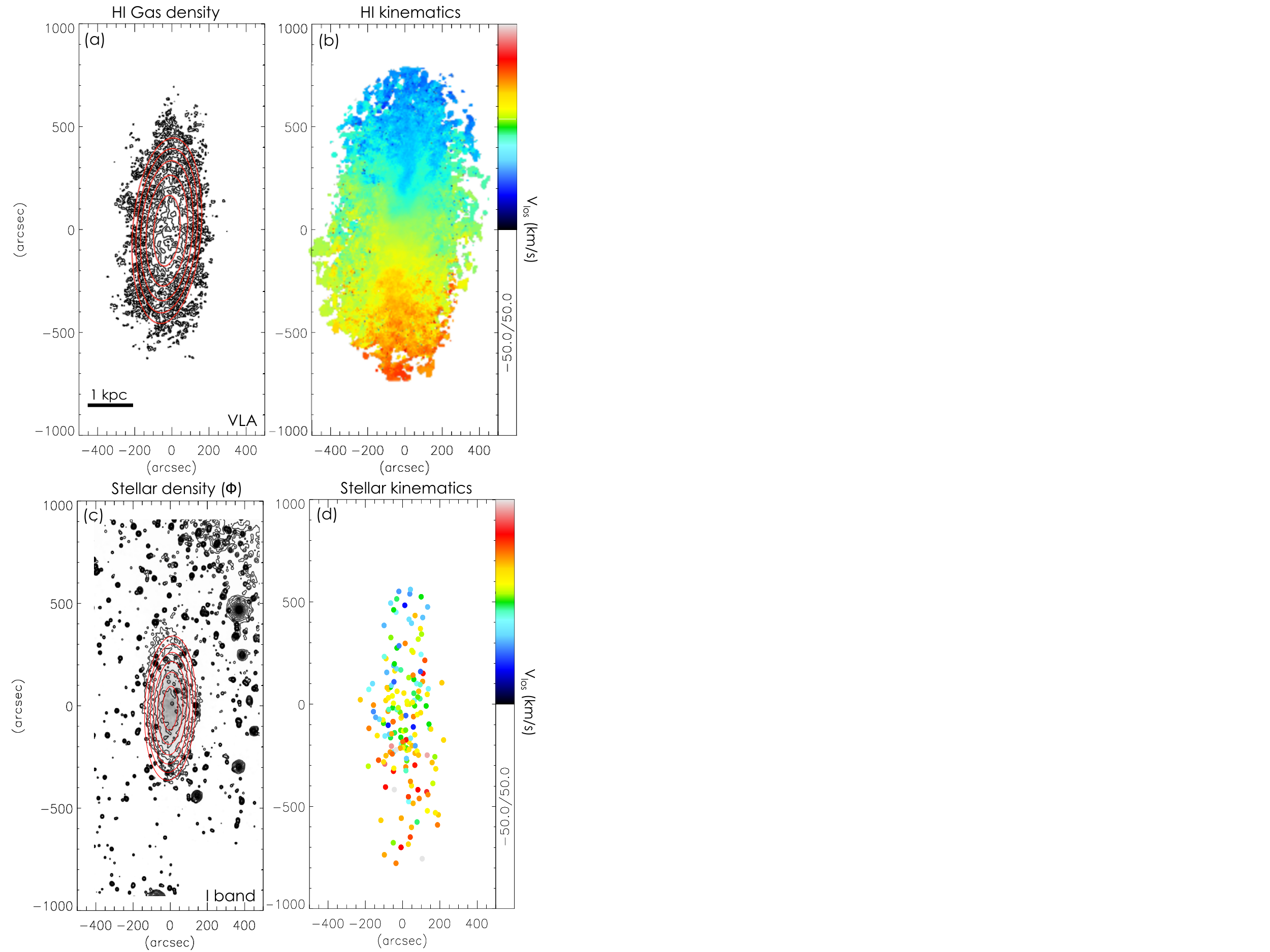}
\caption{Photometric and kinematic data. (a) and (b): HI surface density and velocity maps \citep{ior17}. (c): Greyscale and black contours are the smoothed I band image of WLM. The fitted MGEs are overlaid in red. 
(d) Discrete velocity measurements.}
\label{fig_HIstarobs}
\end{center}
\end{figure}

We utilise a discrete set of line-of-sight velocity measurements from 180 member giant branch stars obtained using FORS2 on VLT and DEIMOS on Keck. The typical uncertainties on velocity are $\delta V\sim 6-9$\,km s$^{-1}$, and the reader is referred to \citet{lea09,lea12,lea13} for details on the data reduction and observations.  This sample has already been cleaned from non-member contaminants on the basis of line-of-sight velocity and position metrics. The position and line of sight velocities of the stellar kinematic members are plotted in Figure \ref{fig_HIstarobs}(e).

\begin{figure*}
\begin{center}
\includegraphics[width=1.0\textwidth, trim = 12 380 8 0, clip = True]{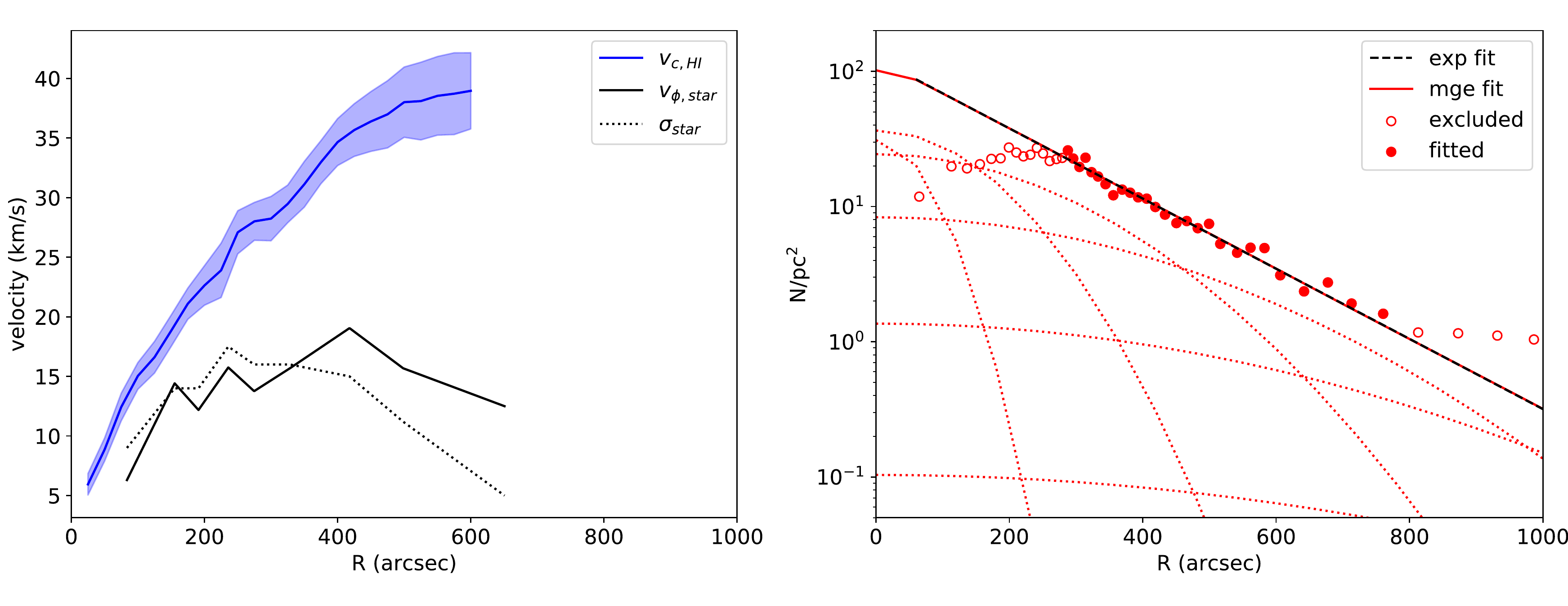}
\caption[HI circular velocities, stellar mean velocity and velocity dispersion and stellar count (RGB stars) radial profiles of WLM]{\textit{Left:} HI Circular velocities derived by \protect\citet{ior17} with the velocity map shown in Figure \protect\ref{fig_HIstarobs}(b) in blue, with 1\,$\sigma$ uncertainties shown by the light blue band. The binned stellar mean velocity ($v_\mathrm{\phi, star}$) and velocity dispersion ($\sigma_\mathrm{star}$) profiles are shown in solid and dotted black lines. \textit{Right:} The RGB star counts are shown as red circles, with the open circles indicating points that are excluded due to crowding and background contamination in the fitting of exponential profile as adopted in \protect\cite{lea13}. The fitted exponential profile is shown in the black dashed line. The individual MGEs fitted to the exponential profile are shown in red dotted lines and the total MGE is shown in a red solid line.}
\label{fig_HIvc}
\end{center}
\end{figure*}

\section{Discrete Jeans Model}\label{sect_model}
Given a total gravitational potential $\Phi$, a velocity anisotropy and an inclination, the Jeans equations \citep{jeans22} specify the projected second velocity moment $V_\mathrm{RMS}^{2} = V_\mathrm{mean}^2 + \sigma^{2}$ of a kinematic tracer of known density, where $V_\mathrm{mean}$ and $\sigma$ are the line-of-sight mean velocity and velocity dispersion. To begin, we assume axisymmetry for WLM and utilise Jeans Axisymmetric Models \citep[JAM,][]{cap08} to solve for the predicted velocity moments. The Jeans equations, under the axisymmetric assumptions, can be written as:

\begin{equation}
\begin{aligned}
&\frac{\partial(R\nu\overline{v_{R}^{2}})}{\partial R}+R\frac{\partial(\nu\overline{v_{R}v_z})}{\partial z}-\nu\overline{v_\mathrm{\phi}^{2}}+R\nu\frac{\partial \Phi}{\partial R}=0,\\
&\frac{\partial(R\nu\overline{v_{R}v_z})}{\partial R}+R\frac{\partial(\nu\overline{v_z^{2}})}{\partial z}+R\nu\frac{\partial \Phi}{\partial z}=0,\\
&\nu\overline{v_\phi^2}(R,z) = \Big(1-\frac{1}{\beta_z}\Big)\Big[R\frac{\partial}{\partial R}\Big(\int_z^{\infty}\nu\frac{\partial\Phi}{\partial z} dz\Big)\\
&+\int_z^{\infty}\nu\frac{\partial\Phi}{\partial z} dz\Big]+R\nu\frac{\partial \Phi}{\partial R}\\
\end{aligned}
\end{equation}
where $\nu(R,z)$ is the surface density of the kinematic tracer and $\Phi(R,z)$ is the axisymmetric gravitational potential. Again, ($v_{R}, v_{z}, v_{\phi}$) are the velocity components in the three dimensions of the cylindrical coordinates ($R, z, \phi$), with $\beta_\mathrm{z}=1-\overline{v_z^{2}}/\overline{v_R^{2}}$ being a velocity anisotropy. Following \cite{cap08}, the velocity ellipsoid is assumed to be aligned with the cylindrical coordinates such that $\overline{v_{R}v_z}=0$.

\subsection{Constructing the potential}\label{sect_pot}
We construct the gravitational potential $\Phi$ with three components, namely, the gaseous component, the stellar component and the dark matter component. Each of the components is parametrised by a set of Multi-Gaussian Expansions (MGEs) \citep{em94}, which deproject and decompose 2D surface densities into superpositions of 3D gaussian components, as is required for our Jeans model. The 2D (projected) surface densities $\Sigma(x',y')$ are first decomposed into gaussians:
\begin{equation}
\Sigma(x',y') = \sum_{k=1}^{N}I_{0,k}\exp\Big[-\frac{1}{2\sigma_k^2}\Big(x'^2+\frac{y'^2}{q_k'^2}\Big)\Big],
\end{equation}
where $I_{0,k}$ is the central density, $\sigma_k$ is the width and $q_k'$ is the observed flattening of each of the gaussian components $k$. The 2D gaussian components are then deprojected to describe the 3D density $\nu(R,z)$:
\begin{equation}
\nu(R,z) = \sum_{k=1}^{N}I_{0,k}\exp\Big[-\frac{1}{2\sigma_k^2}\Big(R^2+\frac{z^2}{q_k^2}\Big)\Big],
\end{equation}
with $q_k=\sqrt{(q_k'^2-\cos^2i)}/\sin i$ describing the intrinsic flattening of each component deprojected given an inclination $i$.  For all the components we adopt the same PA and inclination as the ones derived from the HI velocity map (PA=174$\degree$ and $i$=74$\degree$). Below we provide details on the distributions of the various components.

\subsubsection{Gaseous component}\label{subsect_mgeHI}

We fit MGEs to the HI integrated intensity map using the python code provided by \citet{cap08}. When fitting the MGEs, we fixed the inclination to be 74$\degree$, consistent with the derived inclination from the HI velocity field by \cite{ior17}. Figure \ref{fig_HIstarobs} shows the best-fitted MGEs in red contours overlaid on the HI gas density contours. We normalised the MGEs to the total neutral gas mass of WLM, $1.1\times10^8\,M_\odot$, which is taken from  from the single dish observations of \citet{hunt11}. We apply a correction factor of 1.4 to account for the presence of Helium, this yields a total gas mass of $M_\mathrm{gas, tot}\sim1.54\times10^8M_\odot$. The resultant gaseous MGE parameters, the peak surface density $I_\mathrm{0,gas}$, the width $\sigma_\mathrm{gas}$ and flattening $q_\mathrm{gas}$, of each of the constituent gaussians are presented in Table \ref{tab_mge_HI}. The flattening parameter $q$ is given by the ratio between the short and long axis of each gaussian.
\begin{table}
\begin{tabular*}{0.4\textwidth}{@{\extracolsep{\fill}}ccc}
\hline
I$_\mathrm{0, gas}$ ($M_\odot$\,pc$^{-2}$) & $\sigma_\mathrm{gas}$ ($\arcsec$) & $q_\mathrm{gas}$\\
\hline
\hline
3.775 & 40.58 & 0.28\\
1.854 & 91.71 & 0.30\\
\hline
\end{tabular*}
\caption{Multi-Gaussian Expansions of the gaseous component obtained from HI surface density map.}
\label{tab_mge_HI}
\end{table}

\subsubsection{Stellar component}\label{subsect_mges}
To obtain a smooth stellar distribution, we utilise the I-band photometry which traces evolved stars and avoids the irregular light density profiles of bluer bands caused by the often patchy distribution of young stars. We first smooth the I-band image with a gaussian of width 5$\arcsec$ in order to remove the stochasticity inherent in the nearby resolved systems, and then fit MGEs to the smoothed projected surface density map. The MGEs are then normalised to a total stellar mass. The fitted MGEs are overlaid on top of the I band image in Figure \ref{fig_HIstarobs}(c). The resultant stellar MGEs parameters $I_{0,\star}$, $\sigma_\star$, and $q_\star$, as normalised to $M_\star = 1.1\times10^{7}M_\odot$ \citep{jack07} are presented in Table \ref{tab_mge_s}. Despite the presence of some foreground stars in the image, we find that their presence does not change the MGE fits.  

\begin{table}
\begin{tabular*}{0.4\textwidth}{@{\extracolsep{\fill}}ccc}
\hline
I$_{0, \star}$ ($M_\odot$\,pc$^{-2}$) & $\sigma_\star$ ($\arcsec$) & $q_\star$\\
\hline
\hline
2.750 & 14.74 & 0.50\\
14.72 & 130.8 & 0.41\\
6.239 & 199.0 & 0.42\\
\hline
\end{tabular*}
\caption{Multi-Gaussian Expansion of the smoothed I-band stellar surface brightness profiles used to constrain the stellar mass distribution, normalised to a total stellar mass of $M_\star = 1.1\times10^7\,M_\odot$.}
\label{tab_mge_s}
\end{table}

\subsubsection{Dark matter component}

To model the dark matter contribution to the potential of WLM, we utilise a generalised NFW \citep[gNFW;][]{zhao96} profile to describe our dark matter halo. This has a radial density profile of:
\begin{equation}
\rho(R) = \frac{\rho_\mathrm{s}}{(R/r_\mathrm{s})^\gamma(1+R/r_\mathrm{s})^{3-\gamma}},
\label{eq_dmrho}
\end{equation}
with $\rho_\mathrm{s}$, $r_\mathrm{s}$ and $\gamma$ being the scale density, scale radius and slope of the dark matter profile respectively. To test the influence and degeneracy of non-spherical mass distributions, we also allow the dark matter halo to be axisymmetric with a flattening $q_\mathrm{DM}$ (with $q_\mathrm{DM}=c/a$, where $c$ and $a$ are the axes of the dark matter halo, perpendicular and parallel to the axis of symmetry respectively; and the transformation from cartesian coordinate to $R$ in Eq. \ref{eq_dmrho} is $R = \sqrt{x^2+y^2+(z/q_\mathrm{DM})^2}$). We normalise our DM haloes with  the circular velocities at $r_\mathrm{s}$ such that dark matter haloes with the same ($r_\mathrm{s}$, $\gamma$, $\rho_\mathrm{s}$) but different $q_\mathrm{DM}$ would have the same $V_\mathrm{c}(r_\mathrm{s})$. This normalisation is done so that the parameter $q_\mathrm{DM}$ is only sensitive to the shape of the dark matter halo but not the overall enclosed mass. A dark matter halo parametrised by a particular set of ($r_\mathrm{s}$, $\gamma$, $\rho_\mathrm{s}$ and $q_\mathrm{DM}$) can then be decomposed into MGEs - which together with the gaseous and stellar MGEs, provides a representation of the total gravitational potential of WLM.

\subsection{Surface density of the kinematic tracer}\label{subsect_sdkin}
To obtain the density profile of the kinematic tracer $\nu$, we utilise the discrete giant branch star counts from \cite{lea12}. These star counts are constructed from photometric catalogues which have had a comparable colour and magnitude selection to the spectroscopic sample - thus providing the most representative density distribution for the kinematic tracer population.  The stellar density profile for the kinematic tracers is shown in the right panel of Figure \ref{fig_HIvc} in red circles. The inner flattened number count profile is potentially caused by crowding and we correct for it by fitting first an exponential profile to the star counts beyond the crowded region ($\gtrsim300\arcsec$), as shown in the black line. We then fit MGEs to the black dashed line and we again adopt the same PA and inclination (PA=174$\degree$ and $i$=74$\degree$). The resultant MGE fit is shown by the red solid line and the MGE parameters are listed in Table \ref{tab_mge_rgbexpx}. These MGEs are adopted as the surface density of the kinematic tracer in our models throughout the rest of the paper. Readers interested in how robust our results are with respect to the choice of different profiles can refer to Appendix \ref{app_profile}, where we show the impact of this incompleteness correction on our final results.

\begin{table}
\begin{tabular*}{0.4\textwidth}{@{\extracolsep{\fill}}ccc}
\hline
I$_{0, \star}$ ($M_\odot$\,pc$^{-2}$) & $\sigma_\star$ ($\arcsec$) & $q_\star$\\
\hline
\hline
1.601 &   64.769      &   0.422\\
1.882 &   135.675    &   0.422\\
1.259 &   232.891    &   0.422\\
0.430 &   348.873    &   0.422\\
$7.029\times10^{-2}$ &   476.647    &   0.422\\
$5.344\times10^{-3}$ &   611.309    &   0.422\\
$1.893\times10^{-4}$ &   749.823   &   0.422\\
$2.986\times10^{-6}$  &  893.630   &    0.422\\
$1.233\times10^{-8}$  &  1057.583   &   0.422\\

\hline
\end{tabular*}
\caption{Multi-Gaussian Expansion of the RGB star counts fitted by an exponential profile to measurements within $279\arcsec-813\arcsec$ to avoid bias caused by crowding, normalised to a total stellar mass of $M_\star = 1.1\times10^7\,M_\odot$. Note that since the MGEs are fitted from 1-dimensional star count profiles, we take the outermost $q_\star$ as fitted from the I-band image (Table \ref{tab_mge_s}) as the $q_\star$ for all the MGEs here. }
\label{tab_mge_rgbexpx}
\end{table}

\subsection{Model parameters}\label{subsect_modparam}
The relevant velocity anisotropy for the JAM model is $\beta_\mathrm{z} =1-\overline{v_z^{2}}/\overline{v_R^{2}}$, where $\overline{v_z^{2}}$ and $\overline{v_R^{2}}$ are the second velocity moments along the $z$ and $R$ axes respectively of the cylindrical coordinate system.\footnote{ We note that under the assumptions of the JAM model, the vertical velocity dispersion is intrinsically coupled to the self-gravity of the disk plane, in a quasi-hydrostatic equilibrium, and thus $\beta_{z}$ primarily reflects the vertical mass density distribution of the galaxy - however we show later the insight that other components of the velocity ellipsoid provide on the orbital structure of WLM.} 

Typically the modelled $V_\mathrm{RMS}^\mathrm{mod}$ can be compared directly with the observed $V_\mathrm{RMS}^\mathrm{obs}$ for spatially binned data. In the case of nearby dwarf galaxies, spherical Jeans models have often been applied on the observed $\sigma$ (assuming rotation is negligible) in spatial bins along the major axis of the galaxy \citep[e.g.][]{bat11}. However for fully axisymmetric models, it is more flexible to fit to the discrete stellar kinematic data directly. To do this, we compare the observed line-of-sight velocity $V_{\mathrm{LOS},i}$ of each star $i \in N$, where $N$ is the total number of observed stars, to the probability distribution function of the model line-of-sight velocity $V_\mathrm{LOS, mod}$ at their projected location on the sky-plane ($x_i$, $y_i$). The discrete data are by construction, only providing a single $V_{LOS}$ value, while the relative contributions of $V_\mathrm{mean}^\mathrm{mod}$ and $\sigma^\mathrm{mod}$ to $V_\mathrm{RMS}^\mathrm{mod}$ are not constrained by the Jeans model itself.  We therefore follow \citet{sat80} and \cite{cap08} and introduce $\kappa$ as another free parameter to characterize the amount of rotation the system has relative to an \textit{isotropic rotator}, where $\kappa = \overline{v_\mathrm{\phi}}/\sqrt{\overline{v_\mathrm{\phi}^{2}}-\overline{v_\mathrm{R}^{2}}}$.  As described in \citet{cap08}, $\kappa = 1$ is a rotating system with a symmetric velocity ellipsoid in the $R-\phi$ plane (and spherically isotropic in cases where $\sigma_{z} = \sigma_{R}$), while $\kappa$ approaches 0 when the system angular momentum drops, or the anisotropy increases.  While not a direct analogue for angular momentum, the parameterisation allows for a flexible way to fit the discrete velocity field. Readers interested in the the mathematical procedures with which $\kappa$ decompose $V_\mathrm{RMS}^\mathrm{mod}$ into $V_\mathrm{mean}^\mathrm{mod}$ and $\sigma^\mathrm{mod}$ components can refer to Eq. 35-38 of \cite{cap08}.

Assuming a gaussian velocity probability distribution function, the probability of $V_\mathrm{{LOS}, i}$ at the position of each star $i$ can be written as:

\begin{equation}
\begin{aligned}
\ln P(V_\mathrm{LOS,i}) = &\ln\frac{1}{\sqrt{2\pi((\delta V_\mathrm{{LOS},i})^{2}+(\sigma^\mathrm{mod}(x_i,y_i))^{2}})}\\
&-\frac{1}{2}\frac{(V_\mathrm{LOS,i}-V_{\mathrm{mean}}^\mathrm{mod}(x_i,y_i))^2}{(\delta V_\mathrm{{LOS},i})^{2}+(\sigma^\mathrm{mod}(x_i,y_i))^{2}},
\end{aligned}
\label{eq_pi}
\end{equation}
where $\delta V_\mathrm{{LOS},i}$ is the error of the observed $V_\mathrm{{LOS},i}$.

With the inclination and the position angle fixed ($i$=74$\degree$, $PA$=174$\degree$), the inputs for calculating the likelihood $P(V_\mathrm{los, i})$ through the JAM model with Eq. \ref{eq_pi} are: (1) the gravitational potential $\Phi$ specified by MGEs, (2) the tracer density distribution specified by the stellar MGEs, (3) the velocity anisotropy $\beta_\mathrm{z}$ and (4) the $\kappa$ parameter. The free parameters in constructing $\Phi$ are the total stellar mass $M_\mathrm{\star, tot}$, $q_\mathrm{DM}$, $r_\mathrm{s}$, $\gamma$ and $\rho_\mathrm{s}$. We assume that $\beta_\mathrm{z}$\footnote{We have also ran the models with the Mamon-Lokas profile and found that the fitted anisotropy profile remains constant over the radial range where we have kinematic tracers.} and $\kappa$ are constant with radius. We therefore have seven model parameters: ($M_\mathrm{\star, tot}, M_\mathrm{gas,tot}, \beta_\mathrm{z}, \kappa, q_\mathrm{DM}, r_\mathrm{s}, \gamma, \rho_\mathrm{s}$) (see Table 4). 

\begin{figure*}
\begin{center}
\includegraphics[width=0.95\textwidth,trim=30 0 30 0]{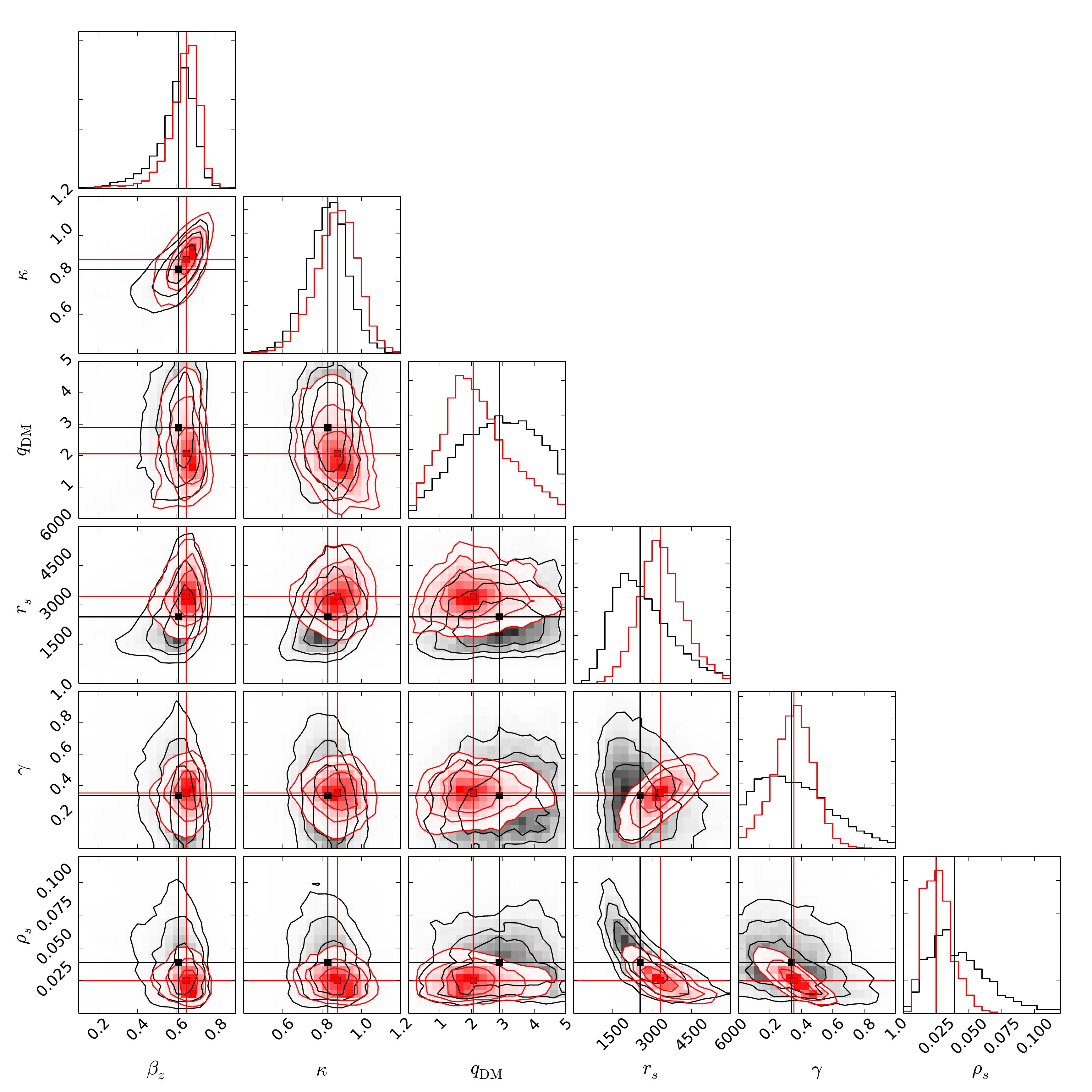}
\caption{Marginalised parameters from the discrete Jeans models: stellar dynamical parameters $\beta_\mathrm{z}$, $\kappa$, and dark matter halo parameters $q_\mathrm{DM}$,  $r_\mathrm{s}$, $\gamma$ and $\rho_\mathrm{s}$. Black contours show the marginalised parameter values with the models using only stellar kinematics, with contour levels 1, 1.5 and 2\,$\sigma$. Red contours show the models run using stellar kinematics and $V_\mathrm{c}$ derived from HI kinematics as a prior.}
\label{fig_emcee_qdmf}
\end{center}
\end{figure*}

\subsection{MCMC sampling}\label{subsect_mcmc}
To obtain marginalised distributions and covariances between the parameters of the most likely models, we sample the likelihood space using the affine-invariant MCMC ensemble sampler implemented in the python package \texttt{EMCEE} \citep{emcee}. We employ 200 walkers, each iterated through 300 steps; the burn-in phase is 100 steps for each walker.

We constrain $M_\mathrm{\star, tot}$ and $M_\mathrm{gas, tot}$ with their observed values, $1.1\times10^7\,M_\odot$ and $1.54\times10^8\,M_\odot$ respectively, through a prior with a normal distribution of width specifying the measurement error of 50$\%$: 
\begin {equation}
\begin{split}
&\ln Pr_1(M_\mathrm{\star, tot}, M_\mathrm{gas, tot})\\
&=\ln\frac{1}{\sqrt{2\pi(0.5\times1.1\times10^{7})^{2}}}-\frac{(M_\mathrm{\star, tot} - 1.1\times10^{7})^{2}}{2\times(0.5\times1.1\times10^{7})^{2}}\\
&+ \ln\frac{1}{\sqrt{2\pi(0.5\times1.54\times10^{8})^{2}}}-\frac{(M_\mathrm{gas, tot} - 1.54\times10^{8})^{2}}{2\times(0.5\times1.54\times10^{8})^{2}}.
\end{split}
\end{equation}

For the other model parameters, we apply a uniform prior; the explored ranges of each of the parameters are listed in Table \ref{tab_prior}.

We run two sets of MCMC processes; one which only uses information from the stellar kinematics (`Stars only') and one with the observed HI $V_\mathrm{c}$ ($V_\mathrm{c, HI}$) as a constrain on the gravitational potential (`Stars + Gas'). In the case for which we include $V_\mathrm{c, HI}$ as a constrain on the gravitational potential, we introduce additionally a second prior term, which evaluate 
\begin {equation}
\begin{split}
&\ln Pr_2(M_\mathrm{\star, tot}, q_\mathrm{DM}, r_\mathrm{s}, \gamma, \rho_\mathrm{s})\\
&=\ln\Big(\Sigma_j\Big(\frac{1}{\sqrt{2\pi(\delta V_\mathrm{c, HI}(R_j))^{2}}}-\frac{(V_\mathrm{c, \Phi}(R_j) - V_\mathrm{c, HI}(R_j))^{2}}{2\times\delta V_\mathrm{c, HI}(R_j)^{2}}\Big)\Big). 
\end{split}
\end{equation}
$\Phi=\Phi(M_\mathrm{\star, tot}, q_\mathrm{DM}, r_\mathrm{s}, \gamma, \rho_\mathrm{s})$ is computed through the MGEs, which gives us $V_\mathrm{c,\Phi}^2(R) = -R(\partial \Phi/\partial R)$. $V_\mathrm{c, \Phi}$ is then evaluated at $R=R_j$, where we have measurements of $V_\mathrm{c, HI}$ from the HI kinematics. Furthermore, we include only combinations of parameters $(M_\mathrm{\star, tot}, q_\mathrm{DM}, r_\mathrm{s}, \gamma, \rho_\mathrm{s})$ which give rise to a $V_\mathrm{c, \Phi}$ that is within 3\,$\sigma$ of $V_\mathrm{c, HI}$, i.e. where $\Sigma_j(\abs{V_\mathrm{c, \Phi}(R_j)-V_\mathrm{c, HI}(R_j)}) < \Sigma_j(3\times\delta V_\mathrm{c, HI}(R_j))$.

The total likelihood for the 180 stars can be written as a sum of the probability and the prior, i.e. $\ln L = \Sigma_{i} (\ln P(V_\mathrm{LOS, i})) + \ln Pr_1$ for the `Stars only' case and $\ln L = \Sigma_{i} (\ln P(V_\mathrm{LOS, i})) + \ln Pr_1 + \ln Pr_2$ for the `Stars + Gas' case.

\begin{table}
\begin{tabular}{@{}ccc}
\hline
parameter & distribution & range\\
\hline
$M_\star$ & normal & $1.1\pm0.56\times10^7\,M_\odot$\\
$M_\mathrm{gas}$ & normal & $1.54\pm0.77\times10^8\,M_\odot$\\
$\beta_z$ & uniform & [-2.0, 1.0]\\
$\kappa$ & uniform & [0.0, 1.5]\\
$q_\mathrm{DM}$ & uniform/fixed & [0.1, 5.0]\\
$r_\mathrm{s}$ & uniform & [500, 10000]\,pc\\
$\gamma$ & uniform & [0.0, 1.0]\\
$\rho_\mathrm{s}$ & uniform & [0.001, 0.15]\,$M_\odot$\,pc$^{-3}$\\
\hline
\end{tabular}
\caption{The adopted priors on each of the model parameters}
\label{tab_prior}
\end{table}

\section{Results}\label{sect_res}

The marginalised model parameters for the set of MCMC runs with free $q_\mathrm{DM}$ are shown in the corner plots in Figure \ref{fig_emcee_qdmf}. Black contours show the dark matter halo and stellar anisotropy parameters constrained from the `Stars only' models, and red contours show the distributions recovered from the `Stars + Gas' models. The corresponding best-fitted parameters and their 1-$\sigma$ uncertainties are listed in Table \ref{tab_emcee}. 

\subsection{DM halo properties}
Both the `Stars only' and the `Stars + Gas' models consistently prefer moderately cored DM profiles, with the posterior distributions showing $\gamma = 0.34^{+0.26}_{-0.21}$ and $\gamma = 0.34^{+0.12}_{-0.13}$ respectively. In Table \ref{tab_gammar}, we list also the derived DM halo density slope (-d$\ln\rho$/d$\ln r$) as a function of fixed radii (expressed as ratio to the half-light radius $r_h$). We show that the DM density slope crosses the cusp-core threshold of 0.5 at $\sim$0.1\,$r_h$ and it is better constrained in the `Stars + Gas` model at all radii by 50- 66\%. A prolate DM halo is preferred in both the `Stars only' and `Stars + Gas' model, with the `Stars + Gas' model indicating a best fit $q_\mathrm{DM}  = 2.1^{+1.3}_{-0.9}$.

While the two models prefer parameters that agree with each other within the uncertainties, it is evident that the dark halo parameters ($r_\mathrm{s}$, $\gamma$, $\rho_\mathrm{s}$) are much better constrained in the `Stars + Gas' models when the HI kinematics are used to jointly constrain the total potential. The uncertainties in the `Stars + Gas' models in $r_\mathrm{s}$, $\gamma$, $\rho_\mathrm{s}$ are smaller than the `Stars only' model by 29$\%$, 48$\%$ and 54$\%$ respectively.   The halo flattening also shows a 15\% reduction in its uncertainty and drives towards more physical prolate values\footnote{Stability analysis for prolate, pressure supported collisionless systems has suggested that axis ratios greater than 5:2 will result in radial orbit instabilities which quickly increase the vertical velocity distribution and reduce the eccentricity \citep{Merritt91}}.

\subsection{Stellar orbital properties}\label{subsect_betar}
Within JAM, the stellar orbital properties are described by $\beta_{z}$ and $\kappa$. $\beta_{z}$ describes the velocity anisotropy and is the best fit models find $\beta_{z} = 0.61^{+0.07}_{-0.12}$ and $0.65^{+0.06}_{-0.09}$ respectively for the `Stars only' and the `Stars + Gas' models. The inclusion of gas kinematics allow a 24\% improvement in the constraint of $\beta_{z}$. It is evident that such an improvement is enabled by breaking the degeneracy between $\beta_z$ and several DM halo parameters such as $q_\mathrm{DM}$, $r_\mathrm{s}$ and $\gamma$. $\kappa$ is constrained to $0.83^{+0.09}_{-0.11}$ and $0.88^{+0.10}_{-0.11}$ respectively for the `Stars only' and the `Stars + Gas' models. The uncertainties of $\kappa$ in both models are similar due to the fact that $\kappa$ is a property that is intrinsic to the stellar kinematical map itself and is not constrained by the Jeans model.

While the anisotropy is described in JAM by $\beta_{z}$, we can study the more informative link with tangential velocity dispersion by computing $\beta_r = 1- (\sigma_\phi^2+\sigma_\theta^2)/ 2\sigma_r^2$. From each of the JAM models we made in the MCMC process, one can compute the individual velocity dispersions in three dimensions: $\sigma_\phi$, $\sigma_{R}$ and $\sigma_\mathrm{z}$ in cylindrical coordinates, which can then be transformed into $\sigma_\phi$, $\sigma_\theta$ and $\sigma_{r}$ in spherical coordinates. Such a calculation can be made following Eqs. 19-23, 32 and 37 from \citet{cap08} with input MGEs describing the gravitational potential $\Phi(R,z)$ and the density profile of the kinematic tracers $\nu(R,z)$, $\beta_z$ and  $\kappa$. Even though we have assumed a radially constant $\beta_z$ and $\kappa$, the radially varying $\Phi$ and $\nu$ render a radially varying $\beta_r$.

\begin{table*}
\begin{tabular}{@{}ccccccc}
\hline
radii ($r_h$) & 0.01 & 0.025 & 0.05 & 0.1 & 0.25 & 0.5\\
\hline
Star only & 0.36$^{+0.25}_{-0.20}$ & 0.38$^{+0.25}_{-0.20}$ & 0.42$^{+0.25}_{-0.20}$ & 0.50$^{+0.26}_{-0.21}$& 0.67$^{+0.30}_{-0.23}$ & 0.91$^{+0.35}_{-0.26}$\\
Star + Gas & 0.38$^{+0.14}_{-0.15}$& 0.40$^{+0.13}_{-0.14}$ & 0.43$^{+0.12}_{-0.13} $& 0.50$^{+0.11}_{-0.11}$& 0.66$^{+0.09}_{-0.08}$ & 0.89$^{+0.12}_{-0.09}$\\
\hline
\end{tabular}
\caption{Slopes of dark matter halo density profile (-d$\ln\rho$/d$\ln r$) as a function of radius (in ratio to the half light radius $r_h$).}
\label{tab_gammar}
\end{table*}

\begin{figure*}
\begin{center}
\includegraphics[width=1.0\textwidth,trim=150 25 150 25, clip = true]{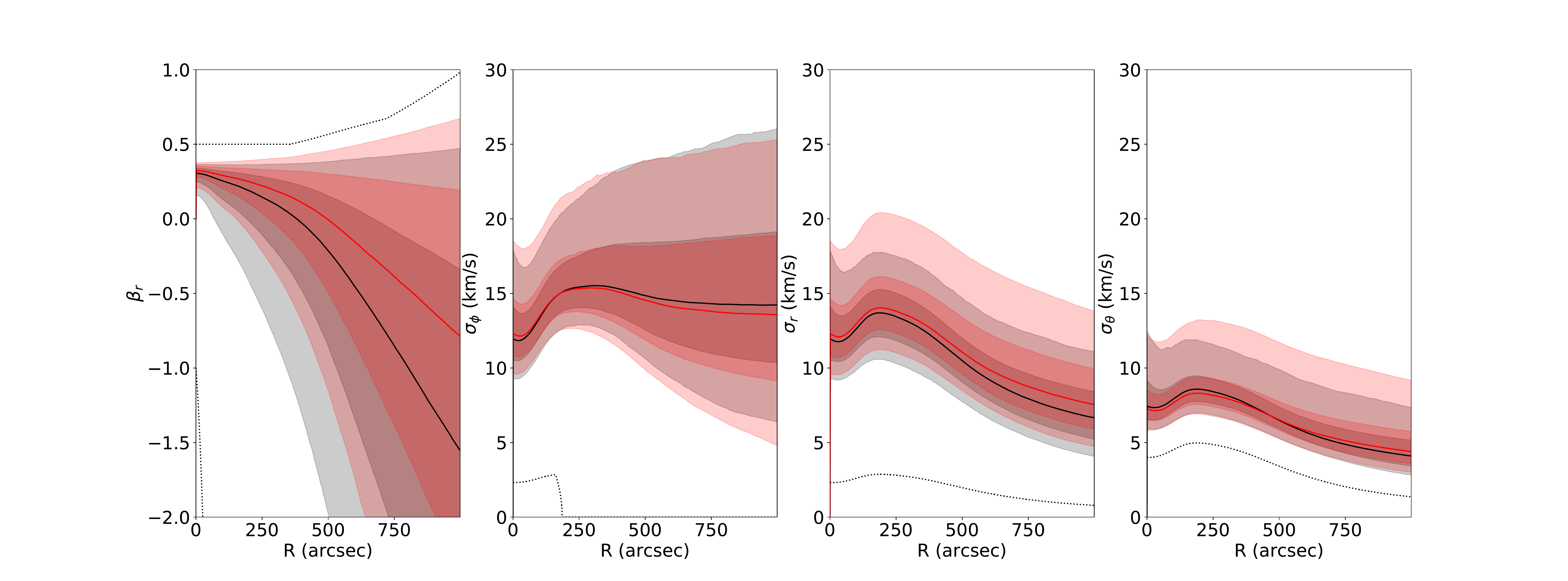}
\caption{From left to right: the derived $\beta_r(R)$, $\sigma_\phi(R)$, $\sigma_r(R)$ and $\sigma_\theta(R)$ at $z=0$ from our dynamical models. The thick black and red lines show the best fitted profile and the bands show the corresponding 1 and 2\,$\sigma$ uncertainties for the `Stars only' and `Stars + Gas' models respectively. The dotted black lines on each panel show the prior on these profiles given our priors for the free parameters as listed in Table \ref{tab_prior}.}
\label{fig_betaphi}
\end{center}
\end{figure*}

Figure \ref{fig_betaphi} shows, from left to right, the derived $\beta_r(R, z=0)$, $\sigma_\phi(R, z=0)$, $\sigma_r(R, z=0)$ and $\sigma_\theta(R, z=0)$ profiles derived from 5000 randomly selected individual MCMC steps in the `Stars only' model in thin lines, with the best-fitted profile indicated by a thick black line and the 1-$\sigma$ uncertainties by a black band. The corresponding profiles for the `Stars + Gas'  models are shown in red. The $\beta_r$ profile transitions from a mildly radial central region to a tangentially anisotropic system in the outer regions. $\beta_r$ goes from $0.32^{+0.03}_{-0.04}$ at $r=0$ to $\beta_r=-0.35^{+0.57}_{-0.90}$ at two half light radii ($2r_\mathrm{h}\sim3300$\,pc) for the `Stars + Gas' models. At $r=2r_h$, the constraint on $\beta_r$ improves by 27\% when incorporating gas kinematics in our model.

\begin{table*}
\begin{tabular}{@{}ccccccccc}
\hline

$\beta_z$ &  $\kappa$ & $\beta_r(r=0)$ & $\beta_r(r=r_h)$ & $\beta_r(r=2r_h)$ & $q_\mathrm{DM}$ & $r_\mathrm{s}$ (pc) & $\gamma$ & $\rho_\mathrm{s} (M_\odot\,\mathrm{pc}^{-3})$  \\
\hline
\multicolumn{9}{c}{Stars only, free $q_\mathrm{DM}$}\\
\hline
0.61$^{+0.07}_{-0.12}$ & 0.83$^{+0.09}_{-0.11}$ & 0.30$^{+0.04}_{-0.06}$ & -0.03$^{+0.24}_{-0.45}$ & -0.81$^{+0.74}_{-1.26}$ & 2.9$^{+1.3}_{-1.3}$ & 2544$^{+1458}_{-948}$ & 0.34$^{+0.26}_{-0.21}$ & 0.039$^{+0.028}_{-0.018}$ \\
\hline
\multicolumn{9}{c}{Stars + Gas, fixed $q_\mathrm{DM}$}\\
\hline
0.49$^{+0.07}_{-0.11}$ & 0.86$^{+0.11}_{-0.11}$ & 0.24$^{+0.4}_{-0.5}$& 0.02$^{+0.22}_{-0.34}$ & -0.45$^{+0.58}_{-0.89}$& 0.5 & 2442$^{+1265}_{-959}$ & 0.25$^{+0.25}_{-0.16}$ & 0.032$^{+0.031}_{-0.015}$\\
0.62$^{+0.06}_{-0.09}$ & 0.85$^{+0.10}_{-0.11}$ & 0.31$^{+0.03}_{-0.05}$ & 0.03$^{+0.23}_{-0.37}$ & -0.64$^{+0.70}_{-1.07}$ & 1.0 & 2142$^{+1041}_{-800}$ & 0.27$^{+0.25}_{-0.17}$ & 0.036$^{+0.031}_{-0.016}$\\
0.63$^{+0.05}_{-0.09}$ & 0.86$^{+0.10}_{-0.11}$ & 0.32$^{+0.03}_{-0.04}$ & 0.06$^{+0.21}_{-0.36}$ & -0.54$^{+0.62}_{-1.15}$ & 1.5 & 2342$^{+1163}_{-834}$ & 0.27$^{+0.26}_{-0.17}$ & 0.038$^{+0.030}_{-0.018}$\\
0.64$^{+0.05}_{-0.07}$ & 0.86$^{+0.10}_{-0.11}$ & 0.31$^{+0.03}_{-0.04} $& 0.05$^{+0.21}_{-0.39}$ & -0.58$^{+0.64}_{-1.14}$ & 2.0 & 3148$^{+1254}_{-919}$ & 0.31$^{+0.16}_{-0.16}$ & 0.028$^{+0.019}_{-0.013}$\\
0.62$^{+0.06}_{-0.10}$ & 0.84$^{+0.09}_{-0.11}$ & 0.31$^{+0.03}_{-0.05}$ & 0.03$^{+0.22}_{-0.38}$ & -0.64$^{+0.69}_{-1.13}$ & 2.5 & 2453$^{+1179}_{-835}$ & 0.28$^{+0.27}_{-0.19}$ & 0.041$^{+0.027}_{-0.017}$\\
0.63$^{+0.06}_{-0.07}$ & 0.84$^{+0.10}_{-0.11}$ & 0.32$^{+0.03}_{-0.05}$ & 0.05$^{+0.22}_{-0.36}$ & -0.59$^{+0.67}_{-1.05}$ & 3.0 & 3136$^{+1006}_{-826}$ & 0.34$^{+0.12}_{-0.15}$ & 0.027$^{+0.016}_{-0.010}$\\
0.64$^{+0.06}_{-0.08}$ & 0.87$^{+0.09}_{-0.11}$ & 0.32$^{+0.03}_{-0.04}$ & 0.09$^{+0.27}_{-0.33}$ & -0.45$^{+0.49}_{-1.01}$ & 3.5 & 2900$^{+1451}_{-1034}$ & 0.27$^{+0.25}_{-0.17}$ & 0.038$^{+0.025}_{-0.016}$\\
0.63$^{0.06}_{-0.09}$& 0.86$^{+0.10}_{-0.11}$ &0.31$^{+0.03}_{-0.05}$ & 0.05$^{+0.22}_{-0.37}$ & -0.58$^{+0.67}_{-1.10}$& 4.0 &  2741$^{+1438}_{-1067}$& 0.31$^{+0.27}_{-0.20}$ & 0.041$^{+0.035}_{-0.020}$ \\
\hline
\multicolumn{9}{c}{Stars + Gas, free $q_\mathrm{DM}$}\\
\hline
0.65$^{+0.06}_{-0.09}$ & 0.88$^{+0.10}_{-0.11}$ & 0.32$^{+0.03}_{-0.04}$ & 0.13$^{+0.18}_{-0.30}$ & -0.35$^{+0.57}_{-0.90}$ & 2.1$^{+1.3}_{-0.9}$ & 3331$^{+926}_{-778}$ & 0.34$^{+0.12}_{-0.13}$ & 0.025$^{+0.012}_{-0.009}$ \\
\hline
\multicolumn{9}{c}{Stars only, fixed $q_\mathrm{DM}$}\\
\hline
0.53$^{+0.06}_{-0.08}$ & 0.92$^{+0.11}_{-0.13}$ & 0.26$^{+0.03}_{-0.04}$ & 0.15$^{+0.15}_{-0.28}$ & -0.08$^{+0.25}_{-0.72}$& 0.5 & 3061$^{+1046}_{-759}$ & 0.27$^{+0.14}_{-0.14}$ & 0.028$^{+0.016}_{-0.012}$\\
0.66$^{+0.05}_{-0.18}$ & 0.89$^{+0.09}_{-0.10}$ & 0.33$^{+0.02}_{-0.04}$ & 0.14$^{+0.18}_{-0.31}$ & -0.29$^{+0.53}_{-0.90}$ & 1.0 & 3406$^{+1247}_{-1014}$ & 0.35$^{+0.13}_{-0.16}$ & 0.023$^{+0.017}_{-0.010}$\\
0.66$^{+0.05}_{-0.06}$ & 0.88$^{+0.09}_{-0.10}$ & 0.33$^{+0.02}_{-0.03}$ & 0.15$^{+0.16}_{-0.15}$ & -0.29$^{+0.50}_{-0.74}$ & 1.5 & 3118$^{+984}_{-855}$ & 0.31$^{+0.14}_{-0.15}$ & 0.028$^{+0.016}_{-0.011}$\\
0.64$^{+0.05}_{-0.07}$ & 0.86$^{+0.10}_{-0.11}$ & 0.32$^{+0.03}_{-0.03}$ & 0.09$^{+0.19}_{-0.28}$ & -0.48$^{+0.53}_{-0.86}$ & 2.0 & 3148$^{+1254}_{-919}$ & 0.31$^{+0.16}_{-0.16}$ & 0.028$^{+0.019}_{-0.013}$\\
0.64$^{+0.05}_{-0.07}$ & 0.86$^{+0.09}_{-0.11}$ & 0.32$^{+0.03}_{-0.04}$ & 0.08$^{+0.18}_{-0.31}$ & -0.51$^{+0.58}_{-0.93}$ & 2.5 & 2989$^{+818}_{-730}$ & 0.31$^{+0.12}_{-0.14}$ & 0.029$^{+0.015}_{-0.010}$\\
0.63$^{+0.06}_{-0.07}$ & 0.84$^{+0.11}_{-0.11}$ & 0.31$^{+0.03}_{-0.04}$ & 0.03$^{+0.22}_{-0.32}$ & -0.68$^{+0.70}_{-0.97}$ & 3.0 & 3136$^{1006}_{-829}$ & 0.34$^{+0.12}_{-0.15}$ & 0.027$^{+0.016}_{-0.010}$\\
0.63$^{+0.05}_{-0.07}$ & 0.83$^{+0.11}_{-0.11}$ & 0.31$^{+0.03}_{-0.03}$ & 0.03$^{+0.22}_{-0.32}$ & -0.69$^{+0.70}_{-0.95}$ & 3.5 & 3113$^{+969}_{-844}$ & 0.33$^{+0.12}_{-0.14}$ & 0.028$^{+0.015}_{-0.010}$\\
0.64$^{+0.05}_{-0.07}$ & 0.86$^{+0.10}_{-0.11}$ & 0.31$^{+0.03}_{-0.04}$ & 0.00$^{+0.21}_{-0.31}$ & -0.80$^{+0.65}_{-0.96}$ & 4.0 & 3148$^{+1254}_{-919}$ & 0.32$^{+0.16}_{-0.16}$ & 0.028$^{+0.019}_{-0.013}$\\
\hline
\end{tabular}
\caption{Best-fitted Jeans model parameters and 1$\sigma$ uncertainties for $q_\mathrm{DM}$ free and $q_\mathrm{DM}$ models at $0.5<q_\mathrm{DM}<5.0$ at 0.5 intervals. }
\label{tab_emcee}
\end{table*}

\subsection{Dependence on $q_\mathrm{DM}$}
While both the `Stars only' and the `Stars + Gas' model prefer a prolate halo, the flattening of the dark matter halo $q_\mathrm{DM}$ has some of the most important correlations with other parameters. While our method does not rely on the thickness of the HI layer to infer halo flattening, and thus should not be biased by assumptions the HI gas opacity \citep{Peters17b}, given the importance of this parameter more examination is warranted.  We would therefore like to understand the degeneracies between the choice of halo flattening and other parameters of interest. To asses this we run models where the DM halo flattening is fixed to values over a grid of $q_\mathrm{DM}$; ($0.25<q_\mathrm{DM}<4.0$, at intervals of 0.25) , in order to evaluate the effect of $q_\mathrm{DM}$ on the stellar dynamical and dark matter properties.

\begin{figure*}
\includegraphics[width=1.0\textwidth,trim=245 0 250 10, clip = true]{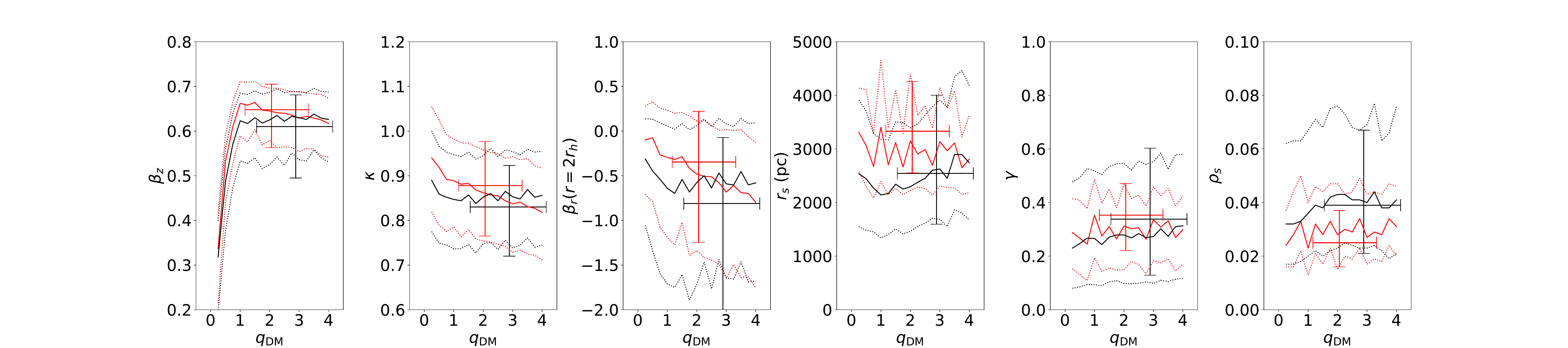}
\caption{The best-fitted (solid line) and 1-$\sigma$ uncertainties (dashed lines) of the parameters constrained from the MCMC process with $q_\mathrm{DM}$ fixed between 0.25 and 4. Models are ran at intervals of 0.25 in $q_\mathrm{DM}$. Black lines show the results from the `Stars only' models and red lines show the results from the `Stars + Gas' models. Error bars show the constrained parameters from the $q_\mathrm{DM}$ free runs.}
\label{fig_allq}
\end{figure*}

The best-fit parameters for these constrained models are plotted as a function of $q_\mathrm{DM}$ in Figure \ref{fig_allq} in solid lines, with the respective 1-$\sigma$ uncertainties indicated by dashed lines. The free parameters are then reported in intervals of $q_\mathrm{DM}=0.5$ in Table \ref{tab_emcee}. Black lines show the parameters constraints from the `Stars only' models and the red lines show the parameters constraints from the `Stars + Gas' models. The best fit parameters from the models where $q_\mathrm{DM}$ is free to vary are also shown by the error bars for reference.

In both the `Stars only' and `Stars + Gas' cases, $\beta_\mathrm{z}$ shows a well known degeneracy with $q_\mathrm{DM}$ at $q_\mathrm{DM}\lesssim1$; a flatter dark matter halo gives a lower $\beta_\mathrm{z}$. Similar degeneracies also exist between $q_\mathrm{DM}$ and $\beta_r$. The derived $\beta_r$ at $r=0$, $r=r_h$ and $r=2\,r_h$ are listed in Table \ref{tab_emcee}. The degeneracies are stronger at large radii ($r\gtrsim r_\mathrm{h}$), with a higher $q_\mathrm{DM}$ corresponding to a lower $\beta_r$ (more tangential anisotropies). Also, the degeneracies between $q_\mathrm{DM}$ and $\beta_r$ extend to much higher $q_\mathrm{DM}$, all the way up to $q_\mathrm{DM}=4$. Curiously, such $\beta_r-q_\mathrm{DM}$ degeneracy is only present in the `Stars + Gas' models but not in the `Stars only' models. The other stellar orbital parameter $\kappa$ also show a degeneracy in the direction of higher $q_\mathrm{DM}$- lower $\kappa$, again such a degeneracy is only present in the `Stars + Gas' models.  

Reassuringly, the inner slope of the DM density profile, $\gamma$ appears robust to the choice of halo shape.  As in the case of the freely varying $q_{DM}$ models, the dark matter parameters, $r_\mathrm{s}$, $\gamma$ and $\rho_\mathrm{s}$, are better constrained on average by 27\%, 39\% and 46\% at all $q_\mathrm{DM}$ when we include $V_\mathrm{c, HI}$ as a constraint.

\section{Discussion}\label{sect_diss}
Using discrete Jeans models, together with circular velocity constraints from the HI gas rotation curve, we have derived tight constraints on the DM halo shape and density profile. Additionally, we derive, for the first time, the stellar velocity anisotropy profile of a dIrr. Below we discuss the implications of our results for modified gravity and dark matter theories, and formation models of dwarf galaxies.

\subsection{WLM's dark matter halo properties in the context of $\Lambda$CDM cosmology}
The halo parameters from our best fit models can be used to reconstruct the three dimensional mass distribution in WLM with high confidence.  Here we examine the inner density profile and flattening of the dark matter halo with respect to simulations of galaxy formation in a $\Lambda$CDM framework.

\subsubsection{Dark matter density profile}
\begin{figure}
\begin{center}
\includegraphics[width=0.6\textwidth, trim=145 100 25 150, clip=true]{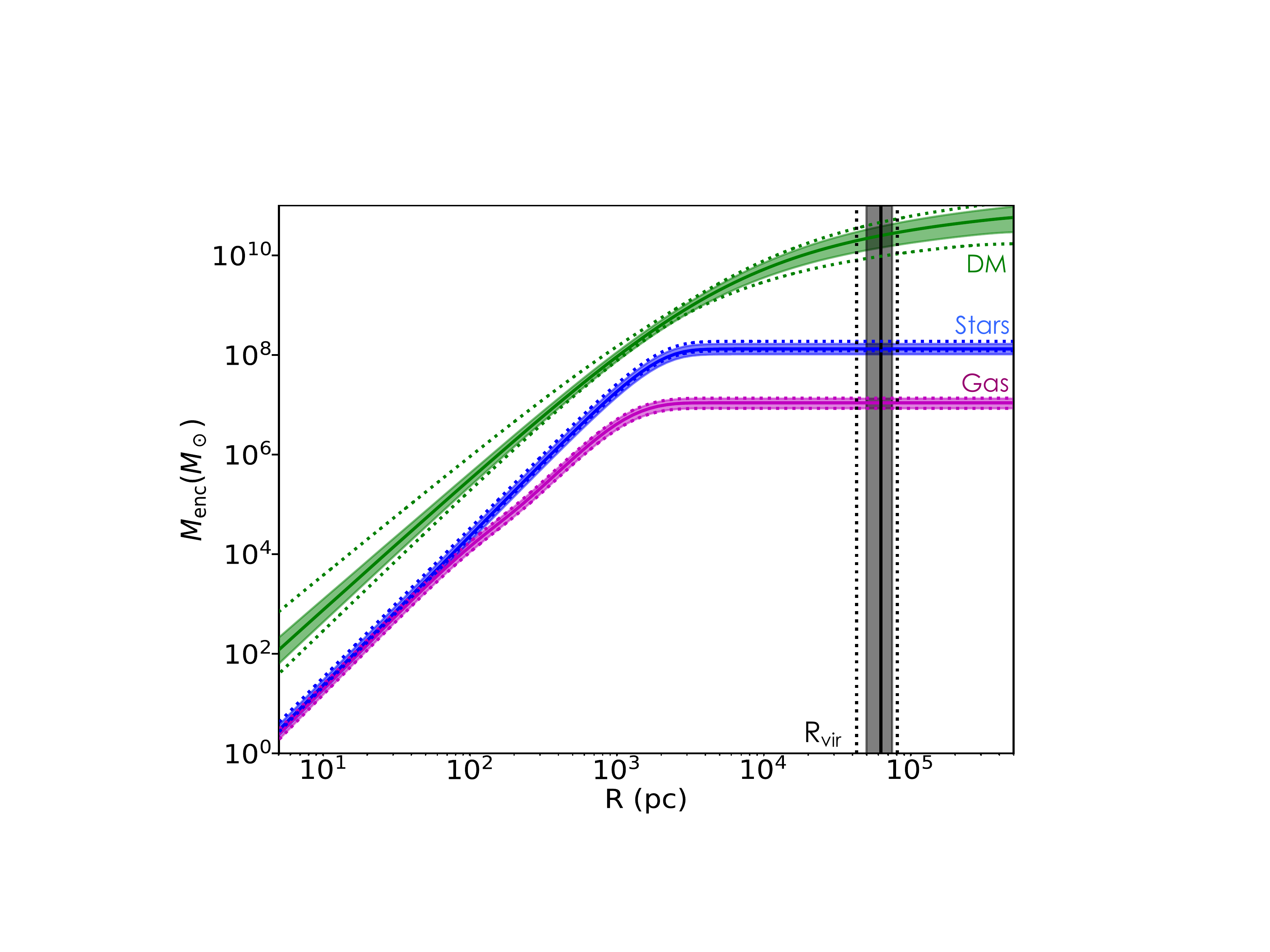}
\caption{Enclosed mass profiles. The stellar, gas and dark matter profile from the best fit`Stars + Gas' model are plotted in magenta, blue and green respectively. Vertical lines indicate the virial radius. Width of the bands give the 1\,$\sigma$ uncertainties. Dotted lines show the corresponding constraints from the `Stars only' dynamical models.}
\label{fig_mprofile}
\end{center}
\end{figure}

Figure \ref{fig_mprofile} shows the dark matter and stellar enclosed mass profiles (within a sphere) derived from our `Stars + Gas' and $q_\mathrm{DM}$ free dynamical model in green and purple respectively. The dark matter virial mass, $M_\mathrm{vir}$\footnote{$M_\mathrm{vir}$ here is defined as the enclosed mass within the virial radius $R_\mathrm{vir}$, where the mass density $\rho(R=R_\mathrm{vir})=200\rho_\mathrm{crit}$ and the critical density $\rho_\mathrm{crit}$ is calculated with a Hubble constant $H=67.1$.}, is constrained to within 2.50$^{+1.75}_{-1.23}\times10^{10}\,M_\odot$ in the `Stars + Gas' model and 2.00$^{+2.89}_{-1.24}\times10^{10}\,M_\odot$ in the `Stars only' model - in good agreement with \citet{lea12}, who used an SIS and NFW fit to the asymmetric-drift-corrected stellar kinematics. 
 
\begin{figure*}
\begin{center}
\includegraphics[width=1.0\textwidth, trim=20 360 175 10]{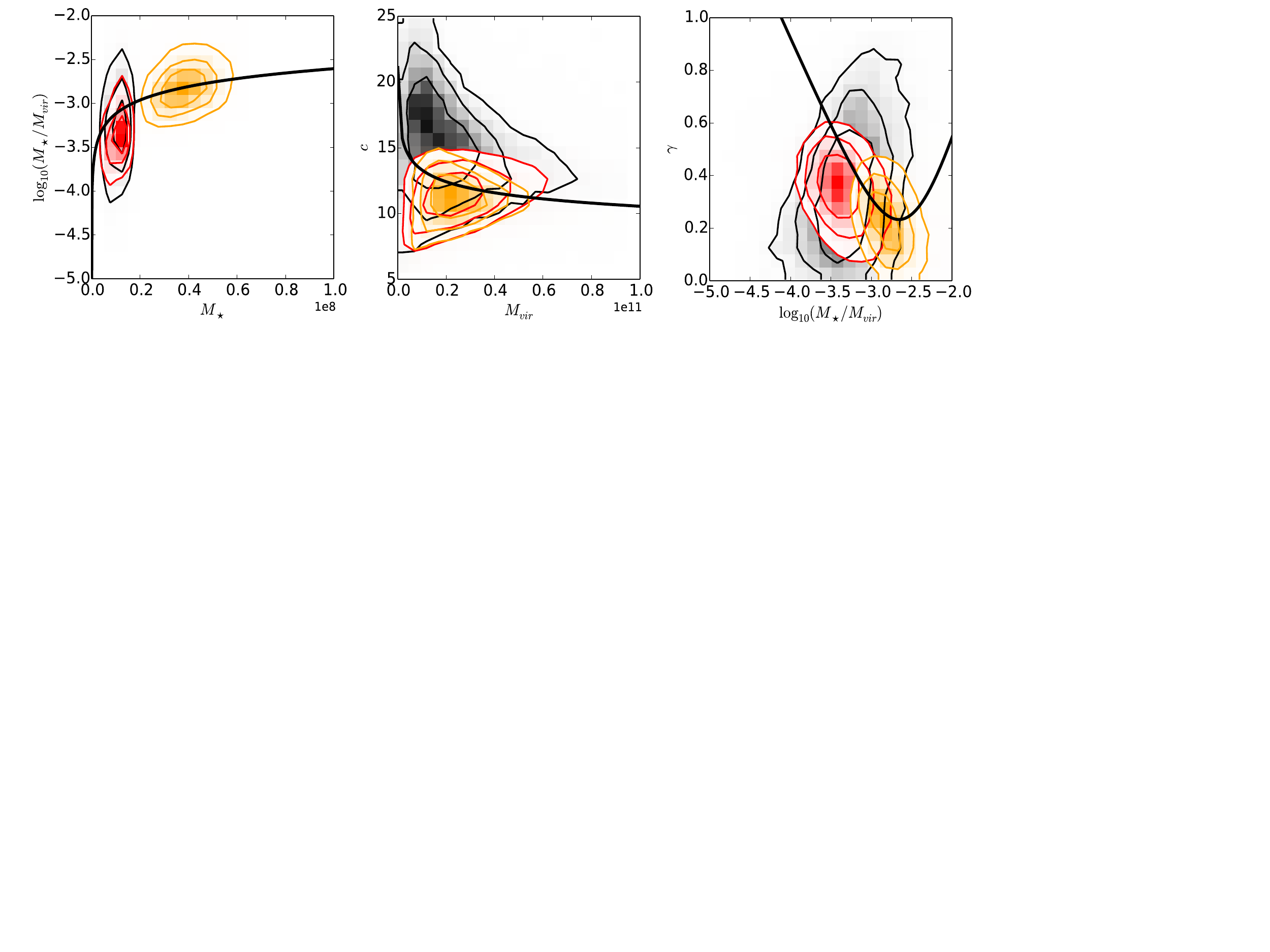}
\caption{Contours showing the constrained values as labeled from models with prior $M_\star=1.1\times10^{7}\,M_\star(\pm50\%)$ `Stars only (black) and `Stars+Gas' (red), and with prior $M_\star=4.3\times10^{7}\,M_\star(\pm50\%)$ `Stars + Gas' in orange. The $M_\star-M_\star/M_{vir}$ relation from \protect\cite{mos10}, the mass-concentration relation from \protect\cite{dut14} and the $M_\star/M_{vit}-\gamma$ relation from \protect\cite{dic14} are shown as thick black lines from left to right panels.}
\label{fig_msm200}
\end{center}
\end{figure*}

The derived stellar to halo mass ratio is therefore $\log_{10}(M_\star/M_\mathrm{vir})=-3.4\pm0.3$, which is slightly higher than the stellar-mass-halo-mass (SMHM) relation found by \citet{mos10} $\log_{10}(M_\star/M_\mathrm{vir})=-3.1\pm0.1$ using the same $M_\star$ value, but consistent within the uncertainties. When we run models with a prior on the stellar mass of $M_\star=4.3\times10^{7}\,M_\odot(\pm50\%)$, a larger value favoured from star formation history studies of WLM \citep{lea17}, we derive a higher $\log_{10}(M_\star/M_\mathrm{vir})=-2.8\pm0.2$.  In  Figure \ref{fig_msm200} we show the $\log_{10}(M_\star/M_\mathrm{vir})$ from the 'Stars only' and `Stars+Gas' models with a prior $M_\star=1.1\times10^{7}\,M_\star(\pm50\%)$ in black and red, and for completeness a `Stars+Gas' model with prior $M_\star=4.3\times10^{7}\,M_\odot(\pm50\%)$ in orange. 

The dark matter halo concentration ($c\equiv r_\mathrm{vir}/r_{-2}$ where $r_\mathrm{vir}$ is the virial radius and $r_{-2}$ the radius at which the logarithmic slope of the dark matter density is $\mathrm{d}\ln\rho_\mathrm{DM}/\mathrm{d}\ln r = -2$) for our best fit models is close to the expected mass-concentration ($M_\mathrm{vir}-c$) relation from dark-matter-only simulations \citep{dut14}. Given our derived $M_\mathrm{vir}$, the $M_\mathrm{vir}-c$ relation found by \citet{dut14} would suggest $c=12.1^{+0.9}_{-0.6}$, consistent within the uncertainties to our inferred halo concentration of $c=11.4\pm1.6$.

Our analysis suggests that WLM has a relatively cored DM density distribution with a best fit to the inner slope of the density profile $\gamma = 0.34 \pm 0.12$.  This value is robust to the recovered DM halo shape ($q_{DM}$), and has an expected correlation with the scale length and normalisation of the dark matter halo, $r_{s}$ and $\rho_{s}$. The central density profile of low mass dwarfs is an important tracer of internal and external evolutionary processes in dwarf galaxies (e.g., \citealt{Zolotov12,Brooks14,Onorbe15}). Using hydrodynamical simulations, \citet{dic14} found that the feedback process which alters the inner slope of dark matter haloes also modifies the final stellar-to-halo-mass ratio ($M_\star/M_\mathrm{vir}$), and a relation between the two was parameterised as:
\begin{equation}
\gamma = -0.06 + \log_{10}[(10^{X+2.56})^{-0.68}+(10^{X+2.56})], 
\label{eq_gammadiC}
\end{equation}
where $X=\log_{10}(M_\star/M_\mathrm{vir})$.

In the mass range of WLM, a higher $M_\star/M_\mathrm{vir}$ would translate to a flatter inner slope (smaller $\gamma$) - as the stellar feedback is proportionally more effective at causing halo expansion due to rapid gas expulsion in the relatively shallow potential well. For our derived $M_\star/M_\mathrm{vir}$, the \citet{dic14} predicts $\gamma=0.5\pm0.2$, consistent within the errors with the $\gamma$ derived from our models of $\gamma = 0.34 \pm 0.12$. If we use the `Stars + Gas' model ran with $M_\star = 4.3\times10^{7}$, the derived value from \cite{dic14}: $\gamma=0.25\pm0.16$ is in excellent agreement with our modelled value: $\gamma=0.23\pm0.12$ (as shown in orange contours in the bottom panel of Figure \ref{fig_msm200}). To compare to the simulations from \cite{read16}, we have also fit our derived dark matter density profile with a cored-NFW profile and found a core size of $r_\mathrm{core}=1257^{+318}_{-269}$\,pc.  In those simulations the typical core size was found to scale with the stellar half mass radius as $r_\mathrm{c} \sim 1.75\,r_\mathrm{h}$.  Our derived core size is slightly smaller than this finding, with the ratio $0.6 \leq r_\mathrm{c}/r_\mathrm{h} \leq 1.0$ for our best fit models. However we note that taking the exponential scale length of the disk \citep[$r_\mathrm{d}=987$\,pc;][]{lea12} gives $0.98\leq r_\mathrm{c}/r_\mathrm{d}\leq1.65$.

In the context of $\Lambda$CDM galaxy formation, WLM appears to have been able to efficiently convert its presumably primordial NFW dark matter cusp into a shallower density profile over a Hubble time of star formation and feedback. This process has occurred, and yet left the system with: an exponential and smoothly distributed intermediate age population \citep{lea12}, no quenched SFH \citep{wei14}, a metallicity distribution function and age-metallicity relation in agreement with a simple leaky box model \citep{lea13}, and a stellar age-velocity dispersion relation consistent with gradual dynamical cooling of the gas \citep{lea17}.  These all suggest that the core-creation process need not always quench the system, nor be catastrophic to the structural, dynamical or chemical properties of the galaxy - at least in this virial mass range.  A more detailed joint analysis of the chemical and kinematic properties may help disentangle whether the core creation process was bursty as expected from feedback scenarios (e.g., \citealt{Kareem17}), or more gradual as in the case of self interacting dark matter.

Previous numerical studies have also explained many of WLM's properties in terms of a feedback based alteration to the underlying NFW profile.  For example, using a set of hydrodynamical simulations for dwarf galaxies, \citet{tey13} were able to reproduce the spatial and dynamical structural properties of WLM, while at the same time transforming the dark matter halo from cusped to core by stellar feedback from bursty star formation. Two WLM-like galaxies with exponential stellar disks of $V/\sigma\sim1$ were also formed in the study by \citet{shen14} from a fully cosmological high-resolution $\Lambda$CDM simulation, again with baryonic feedback playing an important role. The dwarf galaxies from their simulation lie on the observed mass-metallicity relation observed in the Local Group dwarfs, suggesting that the feedback process can operate in a non-destructive fashion for isolated dwarfs.

This provides a counter example to systems such as Ultra-diffuse Galaxies (UDGs), which may acquire their extended structure and old stellar populations partly due to the same feedback processes \citep{dic17}, but with more extreme consequences on the system. Given that some UDGs are estimated to be comparable virial mass to WLM \citep{bea16}, understanding what different conditions during the galaxy's lifetime (e.g. star formation density, environment) lead to such disparate final states is an avenue worth further study.  For example, the resultant decrease in central density and gas concentration may be extremely important for evolutionary changes of dwarf satellites, as demonstrated by \cite{Brooks14}.   Finding present day observational signatures which can trace the rapidity and strength of the potential fluctuations may provide further insight into the timescales, and mechanisms with which the DM core is growing, and can potentially differentiate feedback driven or particle scattering processes (e.g., gas and stellar spatial distributions; \citealt{Mondal18}).  This will be discussed in the subsequent section, however to first order the DM halo density profile we derive is in excellent agreement with the predictions from simulations which incorporate the effect of feedback driven halo expansion in a CDM framework.

\subsubsection{Dark matter halo flattening}
We now turn to the shape (axial ratio) of the dark matter halo inferred from our dynamical models.  Table \ref{tab_emcee} shows that in both the `Stars + Gas' and the `Stars only' models, a prolate dark matter halo is preferred, with $q_\mathrm{DM}\sim2\pm1$ inferred from the `Stars + Gas' model. Pure dark matter $\Lambda$CDM cosmological simulations show that dark matter haloes with our derived $M_\mathrm{vir}$ for WLM have an average short-to-long axis ratio of $\sim0.7$ at the virial radii $r_\mathrm{vir}$ \citep{mac08}. \citet{but16} find similar $q_\mathrm{DM}$ at $r_\mathrm{vir}$ with high-resolution dark matter only simulations. They however extend the analysis towards the inner region and show that over the radii where our stellar kinematics cover ($<5\%\,r_\mathrm{vir}$), dark matter haloes of $M_\mathrm{vir}\sim10^{10}\,M_\odot$ have an even lower average short-to-long axis ratio of $\sim$0.5 and are predominantly prolate.

Those authours used a suite of high-resolution hydrodynamical simulations and showed that while baryonic feedback does not have noticeable effects on $q_\mathrm{DM}$ at the virial radii, it may change $q_\mathrm{DM}$ in the inner region of the halo depending on the $M_\mathrm{vir}$ of the galaxy. The inner region ($<0.12\,r_\mathrm{vir}$) of DM haloes evidently become more spherical for galaxies with $M_\mathrm{vir}>10^{11}\,M_\odot$. For galaxies with $M_\mathrm{vir}$ similar to the one we derived for WLM however, $q_\mathrm{DM}$ does not significantly differ from dark matter only simulations, meaning that a prolate halo with short-to-long axis ratio of $\sim$0.5 is still expected, corresponding to a $q_\mathrm{DM}$ of $\sim$2. This is in excellent agreement with the $q_\mathrm{DM}$ derived from our `Stars + Gas' model. Although a spherical/oblate halo has been ruled out at the 1-$\sigma$ level, such geometries are still possible within the 2-$\sigma$ level. Given the evident $q_\mathrm{DM}-\beta$ (especially $\beta_r$) degeneracies, future proper motion measurements will help us to further constrain the halo geometry.   Similar values consistent with our finding for WLM are seen in the study of \citet{gonz17} with the FIRE simulations of dwarf galaxies.

As we shall see below, the halo shape measurement is a strong prediction of our models, and together with the DM density slope, may offer one of the most powerful lever arms to differentiate baryonic feedback plus CDM scenarios from self-interacting dark matter models.

\subsection{WLM as a test of self-interacting dark matter models and modified gravity}
The simultaneous recovery of a density core and a prolate DM halo is extremely important in understanding the viability of models of non-standard dark matter, e.g., thermal relic, self-interacting (SIDM), Bose-Einstein condensate (BECDM or ``fuzzy'') dark matter.  We have previously seen the good agreement between our observations and predicted values for the DM inner density profile slope and axial ratios in CDM simulations with baryonic feedback.  These models work under the assumption that the DM itself is collisonless and the modifications to the density profile arise indirectly due to stellar feedback rapidly changing the potential well through gas expulsion (c.f., \citealt{pon12}).  

Galaxy formation simulations where the dark matter particle may have a self-interaction cross section, can also produce modifications to the central density profile.  In this case the particle self-interactions, which have a higher rate of occurrence in the denser central regions, result in elastic (or inelastic; \citealt{Vogelsberger19}) scattering of particles (of order one event per particle per Hubble time) and the formation of a density core in the galaxy dark matter distribution \citep{Vogelsberger12,Peter13}.

\begin{figure*}
\begin{center}
\includegraphics[width=0.98\textwidth]{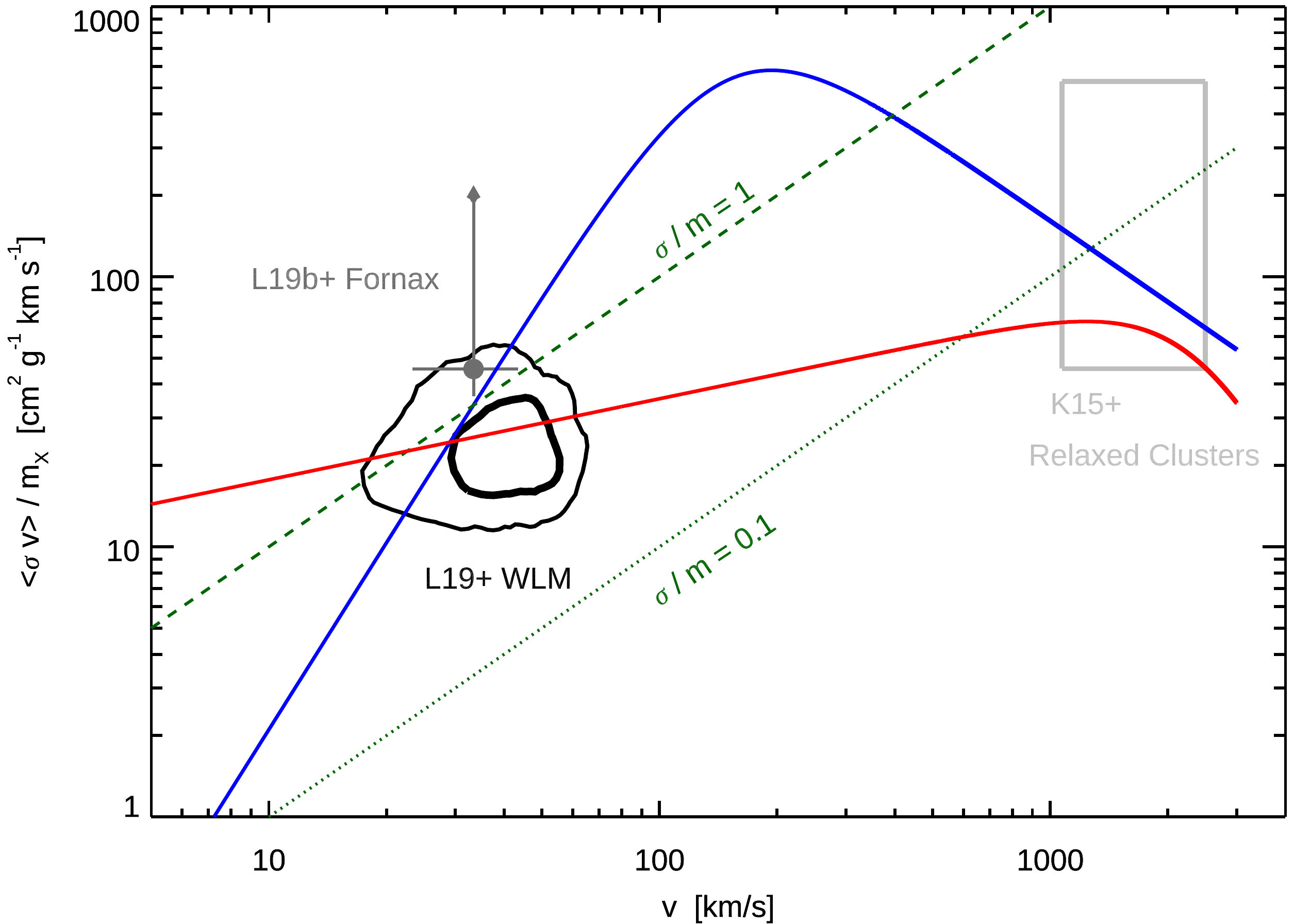}
\caption{Velocity averaged interaction cross sections as a function of characteristic halo velocities.  The self-interacting DM particle cross section necessary to reproduce the density profile of WLM is shown with black contours at the 1- and 2-$\sigma$ levels.  Limits for the Fornax dSph from Leung et al. 2019 are shown in grey.  Dotted and dashed lines show the cross sections for velocity \textit{independent} SIDM models of  $\sigma/m_{X} = 0.1$ and $1$ cm$^{2}$ g$^{-1}$ respectively. 
Two examples of velocity \textit{dependent} SIDM models that are compatible with the dwarf galaxy limits, as well as constraints from galaxy clusters (as marked by the grey box \citet{Kaplinghat16}) are shown in red and blue lines, with parameters indicated. However whether they also preserve aspherical geometries in dwarfs is not yet quantified in simulations.}
\label{fig_wlmsidm}
\end{center}
\end{figure*}

To place our results in the context of such SIDM theories, we compute the model DM density at the core radius $\rho(r_{c})$ using the best fit `Stars + Gas' profile parameters, and derive the likely velocity weighted interaction cross section for SIDM models to produce this cored profile:
\begin{equation}
\frac{\left\langle\sigma v\right\rangle}{m_{X}} = \lbrace\rho(r_{c}) t_{halo}\rbrace^{-1}
\end{equation}
where $m_{X}$ is the mass of the SIDM particle candidate and $t_{halo}$ is the collapse time of the DM halo, here taken to be 13 Gyr.  Figure \ref{fig_wlmsidm} plots the constraints on the cross section using our derived halo properties for WLM.  Also shown are the limits on the same quantity for the Fornax dSph, from Leung et al. 2019, based on modeling of that dwarf galaxy's GC dynamics.  Velocity independent scattering predictions for different SIDM cross sections are shown as green straight lines.  Constraints from high mass galaxy clusters indicate that such velocity independent SIDM models require $\sigma/m_{X} \lesssim 0.1$ cm$^{2}$ g$^{-1}$ (e.g., \citealt{Kaplinghat16}; grey box in Figure \ref{fig_wlmsidm}), which is the dotted green line shown in our figure.  Those studies and others suggest that local dwarf galaxies are more consistent with $\sigma/m_{X} \sim 0.1-10$\,cm$^{2}$ g$^{-1}$. From the `Stars + Gas' model, we derive a $\sigma/m_{X}$ of $0.57^{+0.42}_{-0.20}$\,cm$^{2}$ g$^{-1}$ for WLM.

The mismatch between the required velocity independent cross sections needed for local dwarfs and high mass galaxy clusters has led to velocity \textit{dependent} scattering models to be preferred.  We show three examples as the red, green and blue lines in Figure \ref{fig_wlmsidm}, all of which pass through the combined constraints of WLM and Fornax, but which only the one with the high peak velocity dependence ($v_{max} = 400$ km s$^{-1}$) is also consistent with the cluster measurements of \cite{Kaplinghat16}.  The constraints posed by WLM do not a priori prefer a velocity dependence to the self-interacting DM models - however as we shall see, the \textit{simultaneous finding of a core and a prolate halo may rule out the velocity independent models}, as these are reported to become thermalised and spherically symmetric in their inner regions for the values needed here \citep{Peter13}.

The final core sizes generated from DM scattering can be $\sim 1$ kpc, just as in baryonic feedback + CDM scenarios.  Therefore additional signatures may be needed to differentiate whether a detected DM core is a unique consequence of baryonic feedback, or self-interaction modifications to the DM density profile.  The timescale for the core to form may be longer in SIDM, however this depends on the particular baryonic sub-grid prescriptions adopted (e.g., star formation or feedback injection efficiencies).  For example, \cite{Fry15} showed that the growth rate and final size of the DM core in halos with $V_{max} \leq 30$ km s$^{-1}$ may be largely the same in self interacting dark matter with or without baryonic feedback - though this again depends on the mass range and adopted cross section.  While there could be chemical and/or phase space signatures which may help understand the precise mechanism(s) better, the sparsity of detailed abundances and numbers of observed stars in low mass galaxies makes this a daunting process.  What then may be a potential way to understand whether self-interacting dark matter or feedback scenarios have generated observed cores in dwarf galaxies?

The scattering process that generates a core in self-interaction models may potentially sphericalise the mass distribution, as the interactions are isotropic.  This means that the core formation process in pure self-interaction dark matter models could result in spherical mass distributions in the inner regions of the halos.  The simultaneous quantification of DM density profile slope and axis ratio has unfortunately only been reported as far as we can tell, in simulations of high mass ($M_{vir} \geq 10^{11}$) halos \citep{Peter13}.  In these simulations, halos with $\sigma/m_{X} = 1$\,cm$^{2}$\,g$^{-1}$ which form increasingly cored density distributions (approaching $\gamma \sim 0.4$) become approximately spherical ($c/a \sim 0.9$).  For lower cross sections of $\sigma/m_{X} = 0.1$\,cm$^{2}$\,g$^{-1}$, density profiles slopes of $\gamma = 0.8$ still retain axis ratios of $c/a \sim 0.6$, but these values are not nearly as cored as what we find, and are only reported for halos of $M_{vir} \sim 10^{13-14}\,M_\odot$.   Most importantly, these low values for the cross section are already ruled out on the basis of the WLM DM density profile. 

 Simulations which explore the halo shape of velocity independent SIDM models in the presence of baryons have found that the core creation process can occur with non-spherical final halo shapes in the inner regions \citep{Sameie18}.  However in that case the inner halo progressed towards the axis ratios of the embedded baryonic distribution, which in the case of WLM would be \textit{oblate} with $c/a = 0.4-0.6$ \citep{lea12}.  \cite{Fitts18} simulated dwarf galaxies in our halo mass range with SIDM and baryonic components and found similar behaviour, whereby baryons were the dominant process in altering the DM halo profiles (either indirectly through feedback, or afterwards through contraction) - however there was no reported characterisation of the halo shapes. Velocity dependent SIDM models presented in \cite{Vogelsberger12} show indications that high mass halos can preserve their shapes in the presence of central density modifications, however these simulations were again with MW mass halos.

There is clear need for numerical simulations to quantify the simultaneous evolution of the DM density inner slope and halo shape in the presence of baryons for halos of mass $M_{vir} \sim 10^{10}\,M_\odot$. \citet{rob17} looked at one dwarf in the FIRE simulations which has a stellar mass slightly lower than WLM ($M_\star \sim 10^{7}$). The simulations of SIDM with $\sigma/m = 1$\,cm$^{2}$\,g$^{-1}$ with feedback, and CDM with feedback show the same qualitative behaviour as the Peter et al. 2013 study - SIDM sphericalises the halos as it generates a core in the low mass galaxies also. \textit{ WLM's recovered prolate DM halo with $q_{DM} = 2$, density slope of $\gamma =0.34$ and core of size $r_{c} = 1257$ pc may provide a strong constraint which velocity dependent self-interacting dark matter models need to satisfy}.

Axion mixed DM models or BECDM models also predict a relation between the core size and halo mass - however in this case the core is inherent to the structure formation in these models.  Following \cite{Schive14}, in the case of ultra-light BECDM, the soliton core size is related to the halo virial mass and effective particle mass ($m_{\psi}$) as:
\begin{equation}
r_{c} = 1.5 {\rm kpc}\left(\frac{M_{vir}}{10^{9} M_{\odot}}\right)^{-1/3} m_{\psi}^{-1}
\end{equation} 
For WLM's constraints on the core size and virial mass we find $0.24\times10^{-22} \leq m_{\psi} \leq 1.66\times10^{-22}$ eV/c$^{2}$,  consistent with constraints from large scale structure studies.  Similar to the above SIDM studies more work is needed to quantify the halo axis ratios in low mass halos (with non-negligible baryon fractions), in these or other alternative cosmological models (e.g., ETHOS; \citealt{Vogelsberger16}).

Finally, we comment briefly on the implications of our inferred dark mass distribution on theories of modified gravity such as MOND \citep{Milgrom83}.  WLM is an interesting test case in that it has well defined inclination and measurements of a circular velocity curve from HI kinematics \citep{ior17}, stellar velocity dispersion and anisotropy (\citealt{lea12} and this work) and an intrinsic thickness \citep{lea12}.  Our discrete Jeans model for WLM suggests that there is an extended dark mass distribution around WLM, with a prolate axis ratio of 2:1.  MOND will reproduce the contributions to the observed circular velocity field by altering the acceleration field in the outer regions - however this can only mimic a mass distribution with $q = 0.9$.  WLM is in the deep MOND regime and its extreme isolation means that an external field effect can not be invoked to alleviate discrepancies with MOND predictions in the outer disk.  The prolate dark mass distribution inferred for WLM may represent a significant obstacle for describing the dynamics and structure of this dwarf galaxy with MOND (see also \citealt{Helmi04}).  A follow up paper will present a more detailed discussion and analysis of WLM's stellar structure, dynamics and enclosed mass profile with respect to MOND.

\subsection{Tangential velocity anisotropy in an evolutionary context for dwarf galaxies}
Determining velocity anisotropy in systems with a single type of kinematic tracer has long been assumed to be difficult due to the mass-anisotropy degeneracy inherent to spherical Jeans equations. For a couple of well studied dSphs, authors have used discrete Jeans models, or orbit based Schwarzschild superposition models to better constrain the velocity anisotropy, and found that the anisotropy becomes increasingly more tangential with radius, for both Sculptor \citep{zhu16} and Fornax \citep{kow18}. In subsequent work using proper motions measured from GAIA, \citet{mas18} determined a median radial anisotropy of $\beta_r\sim0.46$ for Sculptor, but only for the inner region  $r\lesssim0.35\,r_\mathrm{h}$.

\begin{figure}
\begin{center}
\includegraphics[width=0.54\textwidth, trim=50 0 0 0]{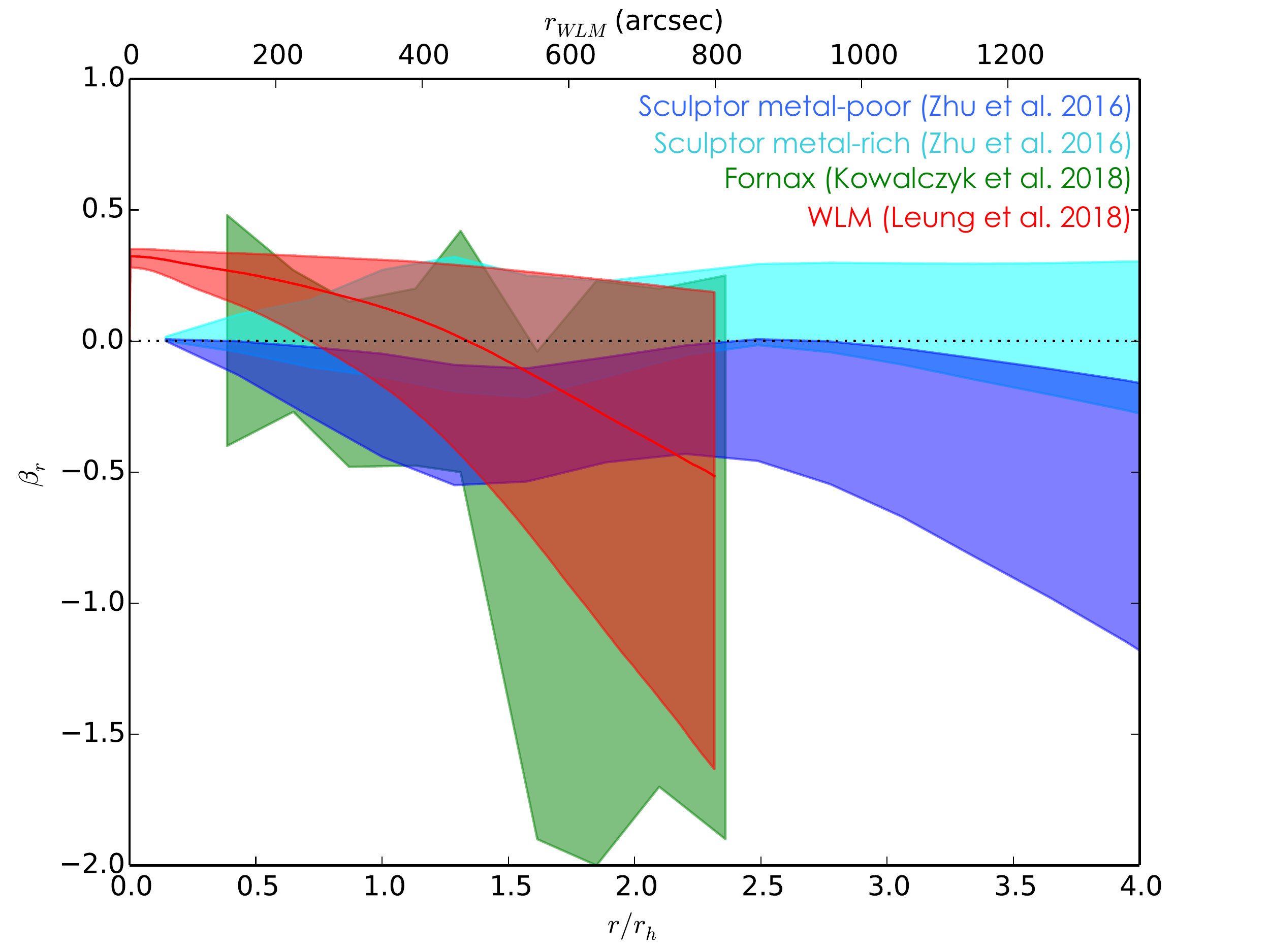}
\caption{Derived WLM $\beta_r$ profile (in red) overlaid on the $\beta_r$ profiles of two dSphs, Sculptor in blue \protect\citep{zhu16} and Fornax in green \protect\citep{kow18}, as an illustration of the similarities in their overall trend. The metal-poor population of Sculptor plotted in cyan has more radial anisotropy but is only dominant in the inner $\sim1.5\,r_h$.}
\label{fig_betaroverlay}
\end{center}
\end{figure}

Interestingly, WLM also demonstrates a mild radial anisotropy in the inner region of $r\lesssim1\,r_\mathrm{h}$, which turns to be tangentially biased towards larger radii ($\beta_r \sim -0.5$). To demonstrate the similarities between the $\beta_r$ profile we obtained from the dIrr WLM and the dSphs, we overlay the $\beta_r$ profiles obtained by \cite{zhu16} for Sculptor (blue) and \citet{kow18} for Fornax (green) on top of the one we obtained from the `Stars + gas' $q_\mathrm{DM}$ free model (red) in Figure \ref{fig_betaroverlay}.  There are clear similarities in all three dwarfs, with the best-fitted anisotropy profile becoming increasingly tangential in the outer regions (albeit the derived uncertainties for both Fornax by \cite{kow18} and WLM by us both allow for slightly radial anisotropy of up to $\beta_r\sim0.2$) .

The interpretation of any anisotropy profile is not straightforward, nor unique.  For example, dissipationless gravitational collapse can lead to an isotropic core, surrounded by an envelope of radially anisotropic orbits \citep{valb82} - however the same configuration is seen to occur in simulations of dwarfs which undergo bar-buckling \citep{may10}. There, bar formation can be triggered by strongly radial anisotropy, before undergoing a bending instability which erases the radial anisotropy (preferentially increasing the vertical velocity dispersion).  In higher mass halos, the reconfiguration of stellar orbits due to minor merging can reproduce the typically radial anisotropic profiles seen for MW mass galaxies, with transient tangential anisotropy appearing due to recent major accretion or flybys of satellites \citep{Loebman18}. 

Alternatively, simulations have shown that tangential anisotropy can be caused by preferential stripping of stars on prograde and radial orbits in a tidal field \citep[e.g.][]{Henon70, Keenan75, bau03, Read06, hur12}. The tangential anisotropy in some dSphs, found especially at large radii ($r\gtrsim r_\mathrm{eff}$, the effective radius), has often been used to support the scenario in which dIrrs are transformed into dSphs via tidal processing. 

The negatively biased $\beta_r$ derived at large radii from our dynamical models for WLM puts the last scenario into question. The velocity anisotropy profiles we find in the dIrr WLM, being nearly isotropic in the center and increasingly tangential towards the outskirts of the galaxy (reaching $\beta_r=-0.35^{+0.57}_{-0.90}$ at $r=2\,r_h$), are very similar to those found in the aforementioned dSphs\footnote{We note here that even though our results do not completely rule out the possibility of a radial anisotropy at large radii within the 1\,$\sigma$ uncertainty, the statistical confidence of tangential bias is comparable with what has been shown in the literature for dSphs and hence the comparison here is justified. Also, we would like to point out that such a tangential bias is not a result our imposed priors; we show in Figure \ref{fig_betaphi} the resulting prior on $\beta_r$ given our input priors the model parameters, as the relevant model parameters all have uniform priors ($\beta_z$, $\kappa$ and DM halo parameters), our prior on $\beta_r$ is also uniform and is allowed to go to highly radial values at large radii.}. WLM is an extremely isolated galaxy ($D_{MW,M31} \sim$1\,Mpc; see Fig. 1 \citet{lea12}), with Local Group barycentric velocity suggesting it has last been in the proximity of a massive neighbour $\sim$11\,Gyrs ago.

WLM's derived $\beta_r$ profile thus provides an environmentally unprocessed baseline for using stellar kinematics to understand the evolutionary similarities or links between dIrrs and dSphs. First of all, the similarity of $\beta_r$ between these dSphs and an isolated dIrr implies that the negative $\beta_r$ seen in dSphs needs not be a result of tidal stripping. The orbital information of both Sculptor and Fornax inferred from proper motion measurements done with GAIA also have weakened the case that they have been tidally stripped \citep{fri18}, as the derived pericentre of these two galaxies are both $\>50$\,kpc and $N$-body simulations on the effects of tides based on observationally motivated orbits on both Sculptor \citep{ior19} and Fornax \citep{bat15}.   

Given the other evidence in its dynamical and chemical evolution for a quiescent existence, it would seem that the tangential anisotropy in this case is either primordial, or imparted through some other mechanism.  Whatever the mechanism to form or impart this anisotropy profile, the similarity between the dIrr and dSphs may also suggest that the transformation from dIrr to dSph is not a violent or dynamical one.  Indeed the stellar kinematics, chemistry and SFHs of some of the \textit{massive} dSphs are becoming increasingly similar to the dIrrs where studies of both are done to comparable depths \citep[e.g.][]{whee17}.  In that case the present day differences may only become extreme where there is significantly early infall, for example for more low mass nearby dSphs - and in other cases perhaps the difference is only quenching of the SF due to gentle ram pressure in the outer halo of the MW's CGM.

If extreme tidal processing is not playing a role in determining the anisotropy profile, we might ask if it it something intrinsic to the formation of galaxies of this mass regime?  Some studies have looked at the relative role of gas pressure support in the initial gas disk of dwarfs \citep{Kaufmann07} or spatial distribution of star formation and stellar populations in dwarf galaxies \citep{Schroyen11}.  However neither study provided quantification of the newly formed stellar anisotropy profiles.   The details of how any aspects of the gas inflow history (e.g \citealt{Keres05}) or turbulence map into 3-D stellar kinematics needs additional study, but may provide help in understanding the similarities in Figure \ref{fig_betaroverlay}.

If the anisotropy at formation is not preserved until present day, the similar profiles for two of the bright classical dSphs and WLM indicate that any evolutionary process which generates tangential anisotropy may need to operate in a generic galaxy of this mass. Such processes could either be connected to dynamical scattering of stars, or the dynamical mixing of gas at the epoch of formation of the surviving stellar populations.

For example, \cite{Christensen16} showed how the re-accretion of gas in the outskirts of MW mass galaxies could introduce flows which have different angular momentum than the local reservoirs.  It is unclear if this would lead to preferential mixing of the newly formed stellar orbits in the tangential direction, or if it could apply in low mass galaxies where there is evidence that a significant amount of the metals in the system may not have been retained or recycled \citep{Kirby11}.

Latent dynamical heating of the stellar orbits in dwarf galaxies may be another mechanism to impart changes in the orbit distribution. 
\cite{lea17} showed that the SFHs of Fornax, Sculptor and WLM were largely consistent with the age-velocity dispersion being a result of dynamical cooling of the ISM as the gas fractions declined over time, however low level scattering of stars was still expected during epochs where the gas and newly formed stellar dispersion was $\leq 5$ km s$^{-1}$.  Individual stars can scatter off of overdensities (e.g., GMCs, spiral arms) in the molecular mid-plane of any galaxy.  

GMC scattering is largely thought to result in both planar and vertical heating and isotropises the stellar velocity ellipsoid, as the stellar disks are much thicker than the molecular gas layers.  Scattering from spiral arms or bars is predominantly planar and so could increase the dispersion in the radial or tangential directions.  However dwarfs of this mass are much too thick and dynamically hot to form spiral arms.  Bar formation has been invoked as an agent important in dwarf galaxy evolution, however the simulations tend to predict either strongly radial (before bar buckling) or vertical (after bar buckling) anisotropies. Also, for dwarf galaxies of mass lower than WLM, bars are not really observed.

Other processes for which increasing evidence is being assembled are the aforementioned feedback driven DM core creation, and dwarf-dwarf mergers.  The non-adiabatic change to the potential induced by the expulsion of gas in the centres of dwarf galaxies is suggested to result in preferentially larger orbit expansion for stars on circular orbits.  If the response of these stars to the largely symmetric change to the potential is a net increase in their orbital radius, then could it be possible that the migrating stars enter final orbits with azimuthal velocities differing from the locally formed stars?  \cite{Kareem17} studied the changes in anisotropy induced by potential fluctuations for dwarf galaxies of this mass, but even though they showed there could be variations, the anisotropy profiles were all significantly radial at all times and locations.  

Mergers have been shown to temporarily induce tangential anisotropy in MW mass galaxies, provided the merging satellite remains coherent in the outskirts \citep{Loebman18}.  However, while there is increasing evidence for dwarf-dwarf mergers in the Local Group, and indeed Fornax (though not recent mergers; Leung et al. 2019), there is no concrete evidence presented in literature for mergers in the other two dwarf galaxies showing tangential anisotropy. A final speculative idea may be that the tangential anisotropy is a \textit{consequence} of the prolate shape of the DM halo.  This will be discussed in a follow up paper.

While the exact cause of the anisotropy profile in WLM and its similarities to those seen in the dSphs is yet unclear, it is clear that the disparate environment posed by WLM offers an important constraint that simulations of isolated field dwarfs (and their potential transformation into dSphs) may want to reproduce.  

WLM is an optimal candidate for the analysis we have presented here as its mass and isolation are both large enough that a significant dynamically cold gaseous component exists.  It is observationally expensive to get stellar kinematics for such objects, but as we illustrate here, the improvement on the recovered dark matter properties are significant.  Among other Local Group dwarf irregulars, few have as well defined HI rotation curves or existing stellar kinematic data sets.  Irregular dwarfs with gas such as IC1613, NGC 6822, Sextans A/B and Pegasus have more chaotic gas kinematic fields or non-optimal inclinations. However Aquarius, Sagittarius dIrr, VV124 may all be possible targets to repeat this type of joint stellar-gaseous dynamical modelling.

\section{Conclusions}\label{sect_con}
We performed Jeans Axisymmetric Models (JAM) on a discrete set of stellar kinematics, consisting of 180 stars, of an isolated dwarf irregular galaxy (dIrr) WLM. The discrete stellar kinematics were obtained using FORS2 on VLT and DEIMOS on Keck, as reported by \citet{lea09, lea12}. Our models incorporated cold HI gas kinematics from \cite{kep07} by introducing the measured circular velocities from HI, $V_\mathrm{c, HI}$, as a prior to the total gravitational potential. We model the dark matter halo with the generalised NFW profile \citep{zhao96}, characterised by the inner slope $\gamma$, the scale radius $r_\mathrm{s}$ and the characteristic density $\rho_\mathrm{s}$. We allow the flattening of the dark matter halo, $q_\mathrm{DM}$, to be a free parameter in our models. The velocity anisotropy is described  by $\beta_\mathrm{z} = 1-\sigma_\mathrm{z}^2/\sigma_\mathrm{R}^2$, which we take to be radially constant for our JAM models. We constrain our model parameters by employing Bayesian statistics. We show that all parameters are better constrained when including $V_\mathrm{c, HI}$ as a prior in our model; the 1\,$\sigma$ uncertainties of the parameters ($\beta_\mathrm{z}$, $q_\mathrm{DM}$, $r_\mathrm{s}$, $\gamma$, $\rho_\mathrm{s}$) improve by 24\%, 15\%, 29\%, 48\% and 54\% respectively.

The dark matter halo is shown to be cored, with $\gamma = 0.34\pm0.12$. Such a cored dark matter halo is robust against variations in the dark matter flattening $q_\mathrm{DM}$ and different $M_\star$ values from the literature. Our inferred $\gamma$ is also consistent with predictions by hydrodynamical CDM simulations, which suggest a relationship between the stellar-to-halo-mass ratio $M_\star/M_\mathrm{halo}$ and the inner slope $\gamma$ of the dark matter halo \citep{dic14}. For our inferred value of $\gamma=0.23\pm0.12$, when adopting $M_\star=4.3\times10^7\,M_\odot$, is in excellent agreement with inner slope inferred by \citet{dic14} of $\gamma=0.25\pm0.16$.

We infer the radial anisotropy profile $\beta_r(r) = 1-(\sigma_\phi^2+\sigma_\theta^2)/ 2\sigma_r^2$ from our JAM models and found that the orbital structure of WLM is characterised by a mildly radially anisotropy core with $\beta_r(r=0) = 0.32^{+0.03}_{-0.04}$ at the centre, which become increasingly tangential and reaches $\beta_r(r=2\,r_h) = -0.35^{+0.57}_{-0.90}$ at 2 half-light radii. This $\beta_r$ profile is very similar to ones obtained from nearby dwarf spheroidal galaxies (dSphs), such as Sculptor and Fornax. While it has been suggested that the tangential anisotropy in dSphs could be caused by preferential tidal stripping of stars on radial orbits, the isolated nature of WLM suggests that the  tangential anisotropy in dwarf galaxies can be of primordial origin and may not be informative on the evolution between dIrrs to dSphs. 

Our model shows that a prolate dark matter halo is preferred in WLM, albeit with relatively high uncertainties: $q_\mathrm{DM} = 2.1^{+1.3}_{-0.9}$. The best-fit value is in good agreement with the dark matter flattening found in $\Lambda$CDM cosmological simulations, both from dark matter only or hydrodynamical simulations, both of which suggest a prolate dark matter halo with $q_\mathrm{DM}\sim2.0$ over the radii covered by our kinematic tracers ($\lesssim5\%$\,$r_\mathrm{vir}$) \citep{but16}. The derived prolate halo suggests challenges to MOND and some self-interacting DM models. These results are currently speculation, however, given the large uncertainties inferred for $q_\mathrm{DM}$. Additionally, we show a $q_\mathrm{DM}-\beta$ degeneracy that extend from $q_\mathrm{DM}=0.5$ to $q_\mathrm{DM}=4.0$ in the `Stars + Gas' models, which provides a window into a better-constrained $q_\mathrm{DM}$ if $\beta$ can be constrained by other means such as proper motion measurements in future spectroscopic observations.

\section*{Data Availiability}
The stellar kinematic data underlying this article are available in the published papers \citep{lea09,lea12}, or by request to the authours.  The stellar photometry are presented in \cite{alan05} and requests for reduced photometry should be directed to those authours. The raw spectra are available via the ESO archive  ({http://archive.eso.org}) and Keck archive \citep{koa}.  The HI observations from \cite{kep07} are available in raw form from the VLA science archive ({https://science.nrao.edu/facilities/vla/archive/index}).

\section*{Acknowledgments}
We thank the anonymous referee for helpful comments which greatly improved the manuscript. We would like to thank G. Iorio for providing the HI circular velocity profiles. GL and GvdV acknowledge support from the German Academic Exchange Service (DAAD) under PPP project ID 57319730. RL acknowledges funding from the Natural Sciences and Engineering Research Council of Canada PDF award and support provided by Sonderforschungsbereich SFB 881 "The MilkyWay System" (subproject A7 and A8) of the German Research Foundation (DFG). G.B. gratefully acknowledges financial support through the grant (AEI/FEDER, UE) AYA2017-89076-P and the MCIU Ram\'on y Cajal Fellowship RYC-2012-11537, as well as by the Ministerio de Ciencia, Innovaci\'on y Universidades (MCIU), through the State Budget and by the Consejer\'\i a de Econom\'\i a, Industria, Comercio y Conocimiento of the Canary Islands Autonomous Community, through the Regional Budget.

\bibliographystyle{mn2e}
\bibliography{examplerefs}   
\appendix
\section{Dependence on the chosen density profile of kinematic tracer}\label{app_profile}
\begin{figure}
\begin{center}
\includegraphics[width=0.5\textwidth, trim = 35 25 450 30, clip = True]{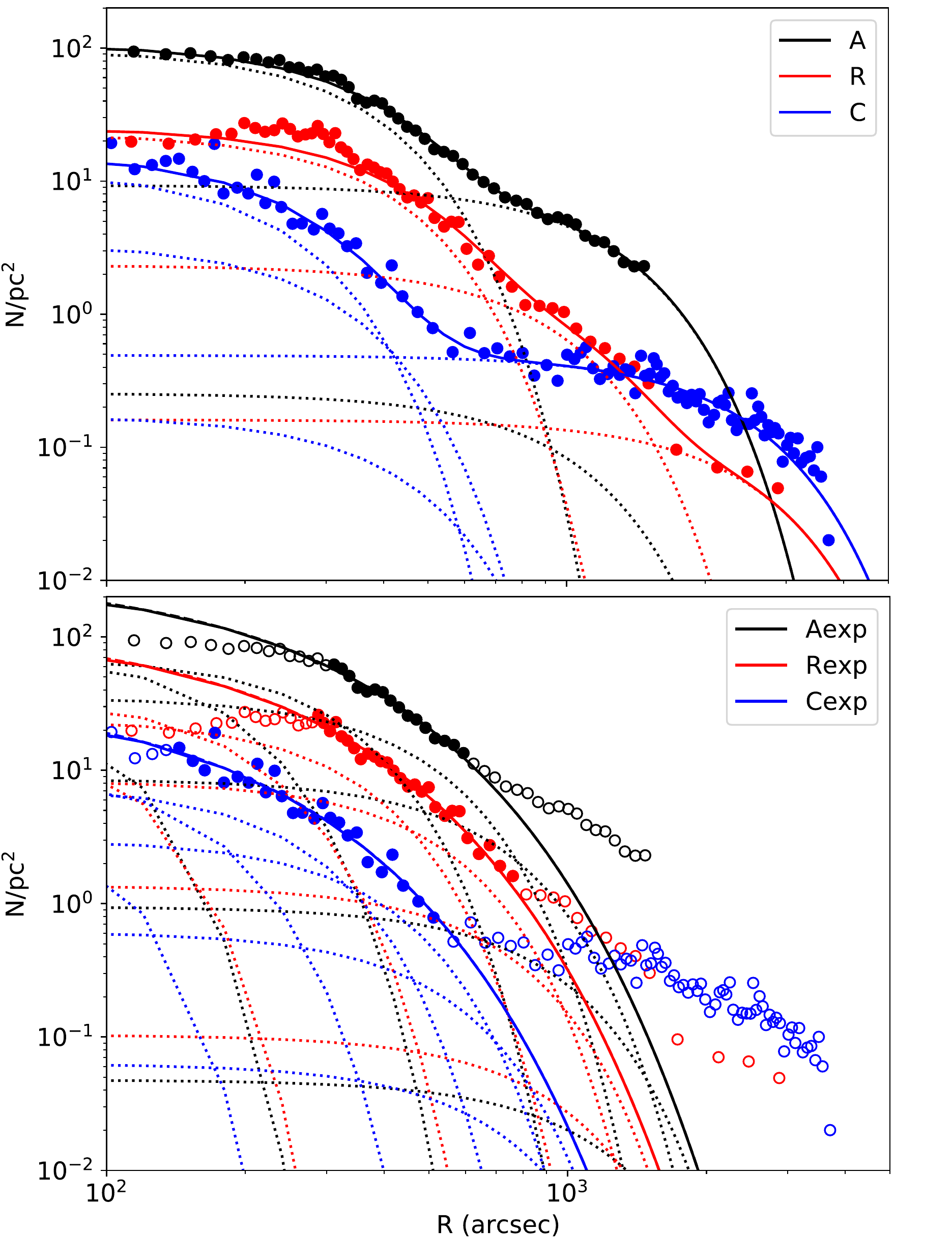}
\caption{Fitted MGEs to RGB stars (red) and C stars (blue). Solid circles show the observed radial profile of the number density of the respective star type. The solids line show the best fitted MGEs and the dotted lines show the individual MGEs. The MGEs fitted from RGB stars are used for both the middle-aged and old populations and the ones from C stars are used for the young population.
\label{fig_binnedmge}}
\end{center}
\end{figure}

\begin{figure*}
\begin{center}
\includegraphics[width=1.01\textwidth,trim=5 310 0 0, clip = true]{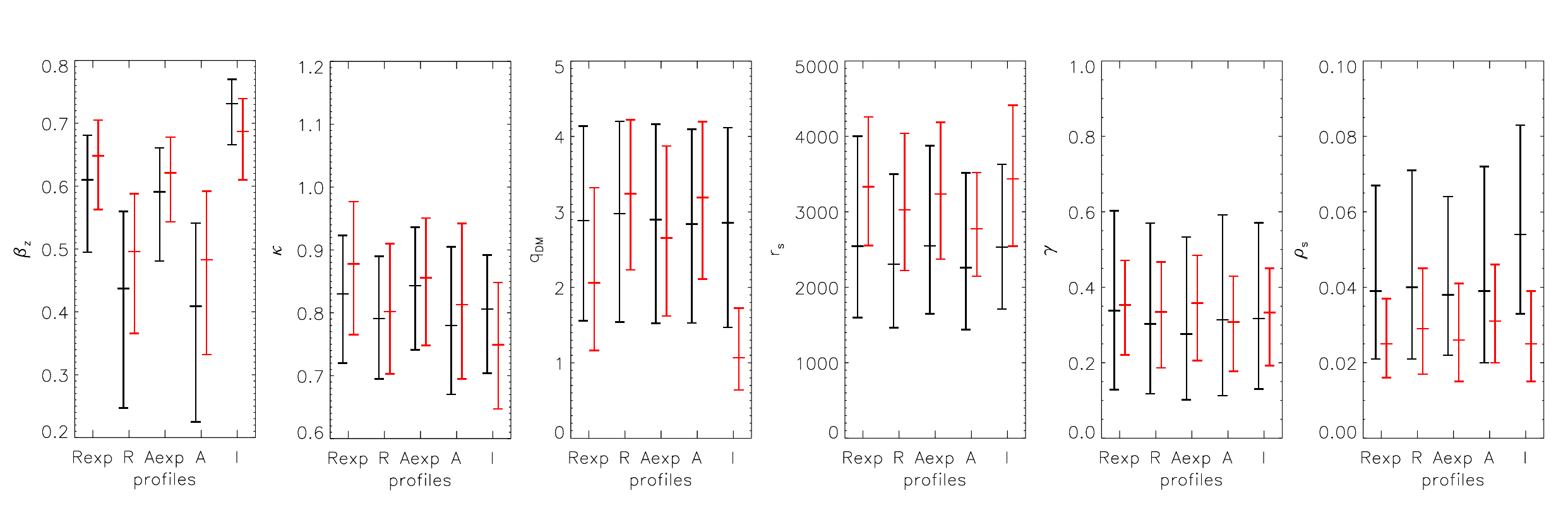}
\caption{Constrained parameters from discrete Jeans models using different density profiles as representation of the kinematic tracer's density profile, with black representing the results from the `Stars only' and red representing the results from the `Stars + Gas' models. The y-axis of each panel shows the constraints of a free parameter in the model, from left to right: velocity anisotropy $\beta_z$, $\kappa$, dark matter halo flattening $q_\mathrm{DM}$, dark matter halo scale radius $r_\mathrm{s}$, inner slope of the dark matter density profile $\gamma$ and the characteristic density $\rho_\mathrm{s}$. The x-axis correspond to the five density profiles that we tested; `Rexp': RGB star counts fitted with an exponential profile excluding the inner region that might be affected by crowding; `R': RGB star counts; `Aexp': total star counts fitted with an exponential profile excluding the inner region; `A': total star counts; and `I': I-band photometry.}
\label{fig_diff_profile}
\end{center}
\end{figure*}

Here we investigate the effects of the chosen input surface density profile of the kinematic tracer. In the main text we have chosen the RGB star counts, fitted with an exponential profile excluding the inner region ($\sim$5$\arcmin$) of the galaxy that might be affected by crowding, to represent the density profile of the kinematic tracer, as shown in Figure \ref{fig_HIvc} and the corresponding MGEs listed in Table \ref{tab_mge_rgbexpx}. We label this profile as `Rexp'. We then rerun the discrete Jeans models on four other density profiles: (1) the uncorrected RGB star counts `R', (2) total star counts with again exponentially corrected profile `Aexp', (3)  uncorrected total star counts `A' and (4) I-band photometry, `I'; the fitted MGE parameters of (1)-(3) are shown in Tables \ref{tab_mge_rgb0x2} to \ref{tab_mge_dsc0x2}, and (4) in Table \ref{tab_mge_s} in the main text. The fitting of the MGEs to the star-count profiles (1) to (3) are shown in Figure \ref{fig_binnedmge}. The best-fitted and 1-$\sigma$ uncertainties of the MCMC parameters constrained from the discrete Jeans model made with each of the profiles are shown in Figure \ref{fig_diff_profile} in black for the `Stars only' case and in red for the `Stars + Gas' case. 

Under all the tested density profiles, a cored dark matter halo with $\gamma<0.5$ is recovered. Furthermore, except for the models ran with I-band photometry as the kinematic tracer's density, a prolate dark matter halo with $q_\mathrm{DM}\gtrsim2$ is preferred. Such a discrepancy is likely caused by the spatial scale at which the density profiles drop off. Its integrated-light nature causes the I-band photometry to drop off at a smaller scale than the other density profiles, which are by nature discrete. The I-band photometry is also shown to have a much smaller spatial coverage than our kinematic tracers (see Figure \ref{fig_HIstarobs}(c) and (d)). The derived $\beta_z$ from the model using `I' as the tracer density profile is also slightly higher than those derived using the other profiles. s kinematics.

\begin{table}
\begin{tabular*}{0.4\textwidth}{@{\extracolsep{\fill}}ccc}
\hline
I$_{0, \star}$ ($M_\odot$\,pc$^{-2}$) & $\sigma_\star$ ($\arcsec$) & $q_\star$\\
\hline
\hline
1.318   &  278.772    &   0.422\\
0.134   &  622.446    &   0.422\\
9.280$\times10^{-3}$   & 1660.687   &    0.422\\
\hline
\end{tabular*}
\caption{Multi-Gaussian Expansion of the RGB star counts (`R'), normalised to a total stellar mass of $M_\star = 1.1\times10^7\,M_\odot$.}
\label{tab_mge_rgb0x2}
\end{table}
\begin{table}
\begin{tabular*}{0.4\textwidth}{@{\extracolsep{\fill}}ccc}
\hline
I$_{0, \star}$ ($M_\odot$\,pc$^{-2}$) & $\sigma_\star$ ($\arcsec$) & $q_\star$\\
\hline
\hline
1.035   &  58.378    &   0.422\\
1.389   &  120.767  &    0.422\\
1.218   &  211.198   &    0.422\\
0.603   &  326.504   &    0.422\\
0.147   &  460.855   &    0.422\\
1.621$\times10^{-2}$  &  607.298    &   0.422\\
8.153$\times10^{-4}$  &   759.947   &    0.422\\
1.838$\times10^{-5}$  &   918.728   &    0.422\\
1.156$\times10^{-7}$  &  1100.052   &    0.422\\
\hline
\end{tabular*}
\caption{Multi-Gaussian Expansion of the exponentially corrected total star counts (`Aexp'), normalised to a total stellar mass of $M_\star = 1.1\times10^7\,M_\odot$.}
\label{tab_mge_dscexpx2}
\end{table}
\begin{table}
\begin{tabular*}{0.4\textwidth}{@{\extracolsep{\fill}}ccc}
\hline
I$_{0, \star}$ ($M_\odot$\,pc$^{-2}$) & $\sigma_\star$ ($\arcsec$) & $q_\star$\\
\hline
\hline
 1.372   &  249.495   &    0.422\\
3.581$\times10^{-2}$    &  668.103   &   0.422\\
 0.131  &   842.278   &    0.422\\
\hline
\end{tabular*}
\caption{Multi-Gaussian Expansion of the total star counts (`A'), normalised to a total stellar mass of $M_\star = 1.1\times10^7\,M_\odot$.}
\label{tab_mge_dsc0x2}
\end{table}

\section{Comparison to spherical Jeans Model}
\begin{figure}
\begin{center}
\includegraphics[width=0.55\textwidth,trim=65 0 390 0, clip = true]{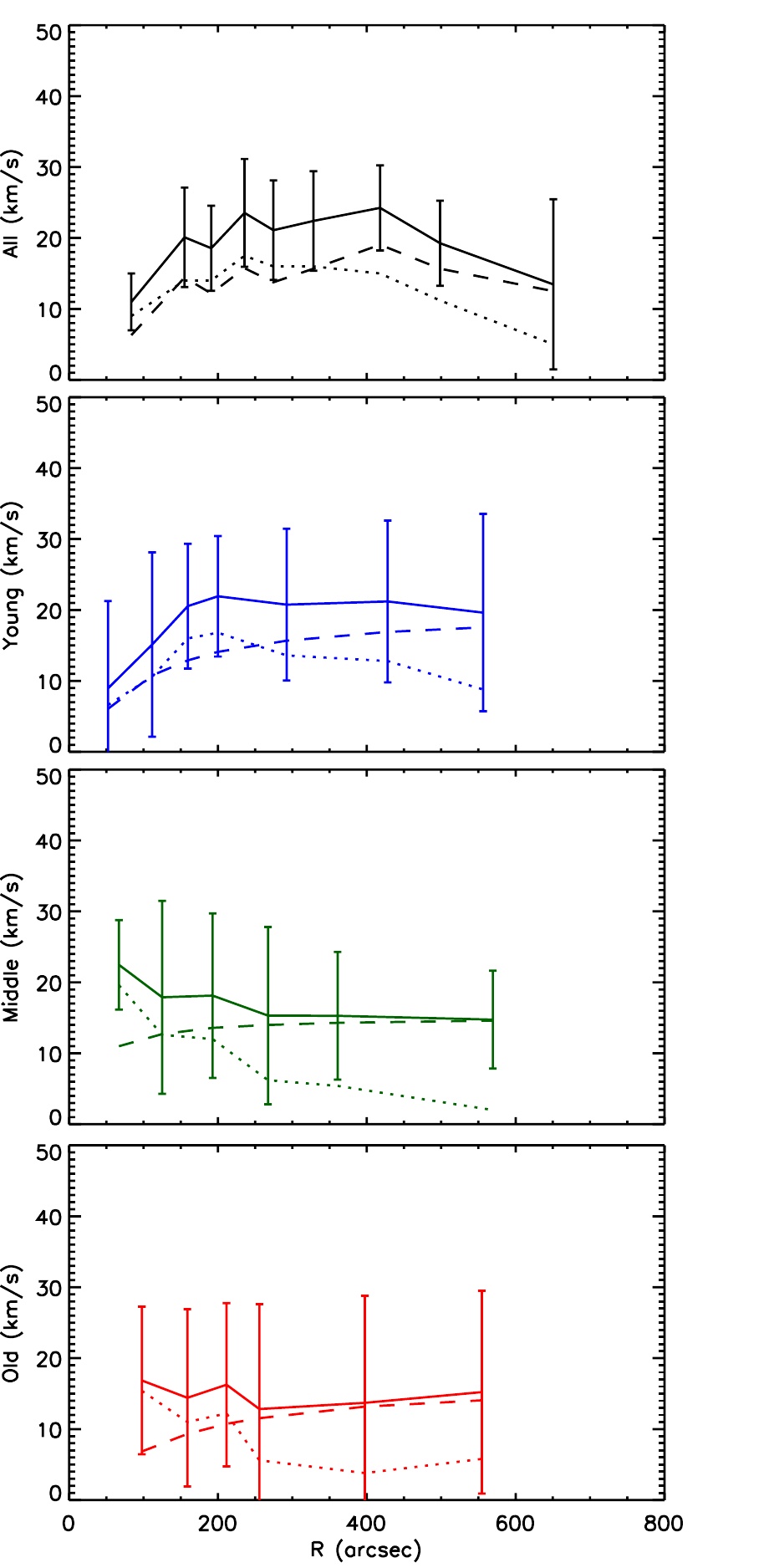}
\caption{Binned stellar mean velocity and velocity dispersion. Top panel: The binned mean velocity $V_\mathrm{mean}$(dashed line) and velocity dispersion $\sigma$ (dotted line) of all the stars in our discrete sample. The solid line show the second velocity moment $V_\mathrm{RMS} = \sqrt{V_\mathrm{mean}^2+\sigma^2}$ as an input to the Jeans model. The binned $V_\mathrm{mean}$, $\sigma$ and $V_\mathrm{RMS}$ profiles of the young, middle and old populations are shown in blue, green and red respectively.}
\label{fig_binnedv}
\end{center}
\end{figure}

Here we compare the dynamical and dark matter parameters as constrained from our JAM model with spherical Jeans model that are commonly used for dwarf galaxies. We use radially binned mean velocity ($V_\mathrm{mean}$) and velocity dispersion ($\sigma$) of our discrete kinematics and the spherical Jeans equation, implemented using the publicly available code by \citet{cap08}. The $V_\mathrm{mean}$ and $\sigma$ profiles are shown in dashed and dotted lines respectively on the top panel of Figure \ref{fig_binnedv}, the corresponding observed second moment $V_\mathrm{RMS} = \sqrt{V_\mathrm{mean}^2+\sigma^2}$ and the error bars are plotted in solid lines. The gaseous and stellar MGEs used are the same as the ones listed in Table \ref{tab_mge_HI} and Table \ref{tab_mge_s}, but with $q=1$ and renormalised to the total stellar and gaseous masses respectively. The dark matter haloes are parametrised with a gNFW profile. 

\begin{figure}
\begin{center}
\includegraphics[width=0.5\textwidth,trim=0 0 0 0, clip = true]{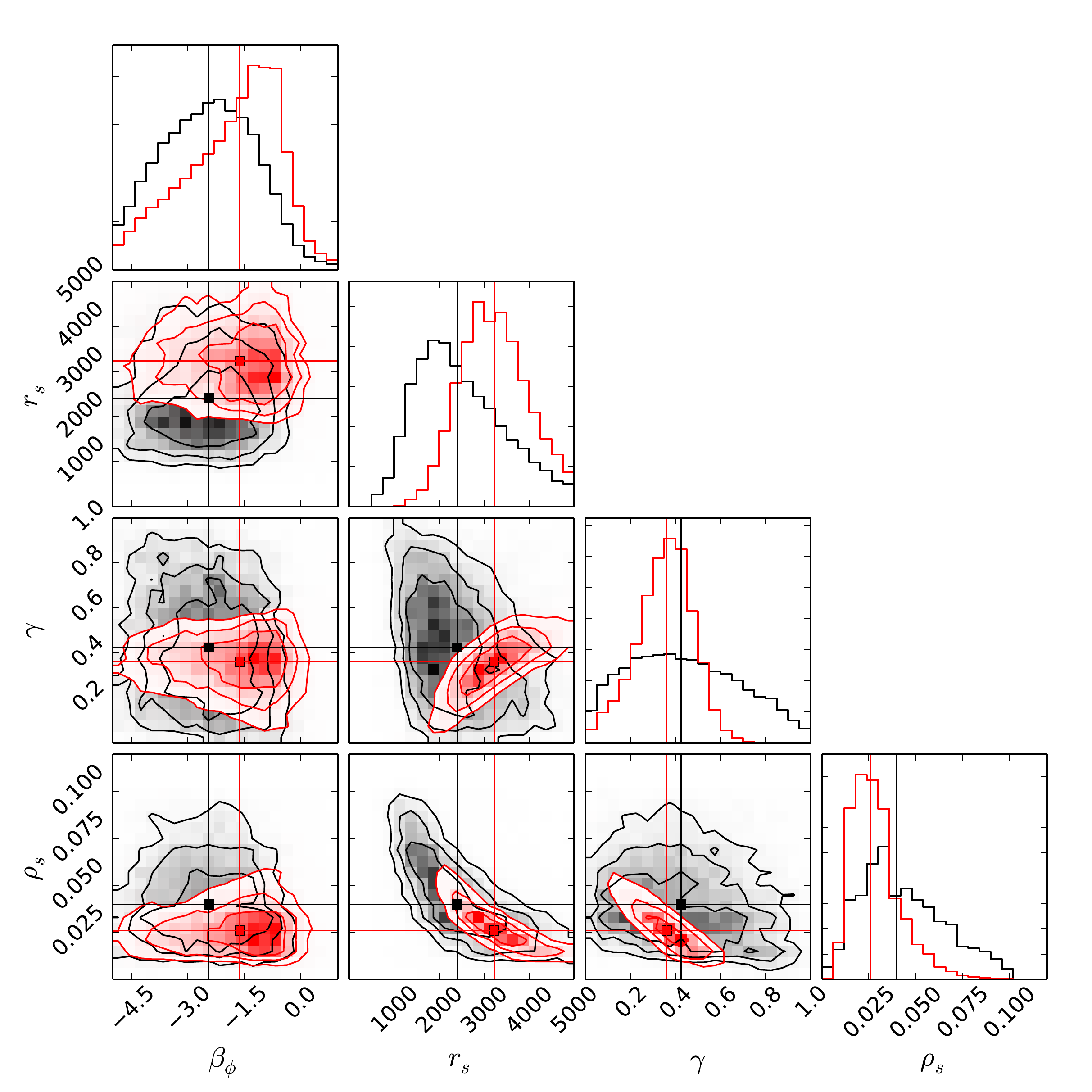}
\caption{Marginalised parameters from the spherical Jeans models, adopting the `Rexp' profile as the surface density profile of the kinematic tracer: the dynamical parameters $\beta_\phi$, and the dark matter parameters$r_\mathrm{s}$, $\gamma$ and $\rho_\mathrm{s}$. Black contours show the marginalised parameter values with Jeans models performed on stellar kinematics only. Red contours show the ones constrained by using $V_\mathrm{c}$ derived from HI kinematics as a prior.}
\label{fig_emcee_sph}
\end{center}
\end{figure}

We again use MCMC to fit the spherical Jeans models to the data, adopting the `Rexp' as the density profile of the kinematic tracer with $q=1$ for all MGEs. The number of walkers, steps and burn-in are the same as the ones we adopt in the axissymmetric case. Since we are using binned data, there is no need to specify $\kappa$. The relevant velocity anisotropy in the Jeans model is $\beta_\phi = \beta_\theta = 1-\sigma_\phi^2/\sigma_r^2. $\footnote{Under spherical symmetry, this would correspond to the radial anisotropy parameter defined in Section \ref{subsect_betar}: $\beta_r = \beta_\phi = \beta_\theta$.} The free parameters are therefore $M_\star$, $\beta_\phi$, $r_\mathrm{s}$, $\gamma$ and $\rho_\mathrm{s}$, we assume $\beta_\phi$ to be constant. We again perform two sets of models, one with constrains from $V_\mathrm{c, HI}$ and one without. The constrained parameters are plotted in Figure \ref{fig_emcee_sph}, in black are the models from the `Stars only' runs and in red the models from the `Stars + Gas' runs.

Just like in the axissymmetric models, the dark matter parameters are much better constrained when we include $V_\mathrm{c, HI}$ as a constraint on the total gravitational potential. The result from the axissymmetric model of a cored dark matter halo remains robust under the spherical Jeans model, which derives a $\gamma$ of $0.37^{+0.11}_{-0.14}$ in the `Stars + Gas' case. Although $\beta_\phi$ is poorly constrained in both the `Stars only' and the `Stars + Gas' cases, it is confirmed here that the stars have a tangential velocity anisotropy, with $\beta_\phi$($=\beta_r$) being highly negative ($-1.67^{+1.03}_{-1.66}$ in the `Stars + Gas' case), just as we find from our discrete JAM models. There is no significant improvement in the constraint on stellar velocity anisotropy by including $V_\mathrm{c, HI}$, reaffirming our interpretation that the improvement of the constraint of $\beta_z$ in the axisymmetric models when including $V_\mathrm{c, HI}$ comes mainly from breaking the $q_\mathrm{DM}-\beta$ degeneracy.

\begin{table}
\begin{tabular*}{0.4\textwidth}{@{\extracolsep{\fill}}ccc}
\hline
I$_{0, \star}$ ($M_\odot$\,pc$^{-2}$) & $\sigma_\star$ ($\arcsec$) & $q_\star$\\
\hline
\hline
   2.426  &    54.426   &    0.422\\
   2.801  &   107.155  &     0.422\\
   1.980  &  178.505   &    0.422\\
   0.774  &   263.719  &     0.422\\
   0.157  &   358.019  &     0.422\\
   1.611$\times10^{-1}$ &    458.012   &    0.422\\
   8.175$\times10^{-4}$ &    561.771   &    0.422\\
   1.982$\times10^{-5}$ &    668.999   &    0.422\\
   1.986$\times10^{-7}$ &    782.365   &    0.422\\
   4.608$\times10^{-10}$  &   912.600 &      0.422\\
\hline
\end{tabular*}
\caption{Multi-Gaussian Expansion of the exponentially corrected C star counts (`Cexp'), normalised to a total stellar mass of $M_\star = 1.1\times10^7\,M_\odot$.}
\label{tab_mge_cexpx2}
\end{table}

\subsection{Multi-population spherical Jeans models}
It has been shown that the stellar velocity anisotropy depends on their metallicity, and by separating the stars into a metal-rich and a metal-poor population one can obtain a better constrain on the velocity anisotropy \citep[e.g.][]{bat06, bat11}. We test here whether we can obtain an even better constrain by adding the $V_\mathrm{c, HI}$ constrain to the multi-population models.

\citet{lea09} have shown that the metal-rich and metal-poor populations in WLM share similar spatial distributions. Here we instead separate the stars into three populations by their ages and characterise their spatial distributions with density profiles from C and RGB stars. The C stars profile is used for the young population ($<$2\,Gyr), the RGB stars profile used for the middle (2-10\,Gyr) and old populations ($>$10\,Gyr). We adopt here the `Rexp' and `Cexp' (an exponential fit to the C stars profile neglecting the inner 2$\arcmin$ for which the fitted MGE parameters are listed in Table \ref{tab_mge_cexpx2}) profiles which avoid issues with over-crowding of stars at the center of the galaxy. We then fit MGEs to the derived exponential profiles. The MGE fittings are shown in Figure \ref{fig_binnedmge} in red for the RGB stars and blue for the C stars. The $V_\mathrm{mean}$, $\sigma$ and $V_\mathrm{RMS}$ for the young, middle and old populations are shown in Figure \ref{fig_binnedv} in blue, green and red respectively. The free parameters here are the velocity anisotropies for the young, middle and old age populations: $\beta_{\phi,y}$, $\beta_{\phi,m}$ and $\beta_{\phi,o}$, and the dark matter parameters $\gamma$, $r_\mathrm{s}$ and $\rho_\mathrm{s}$.

\begin{figure*}
\begin{center}
\includegraphics[width=0.8\textwidth,trim=20 0 20 0]{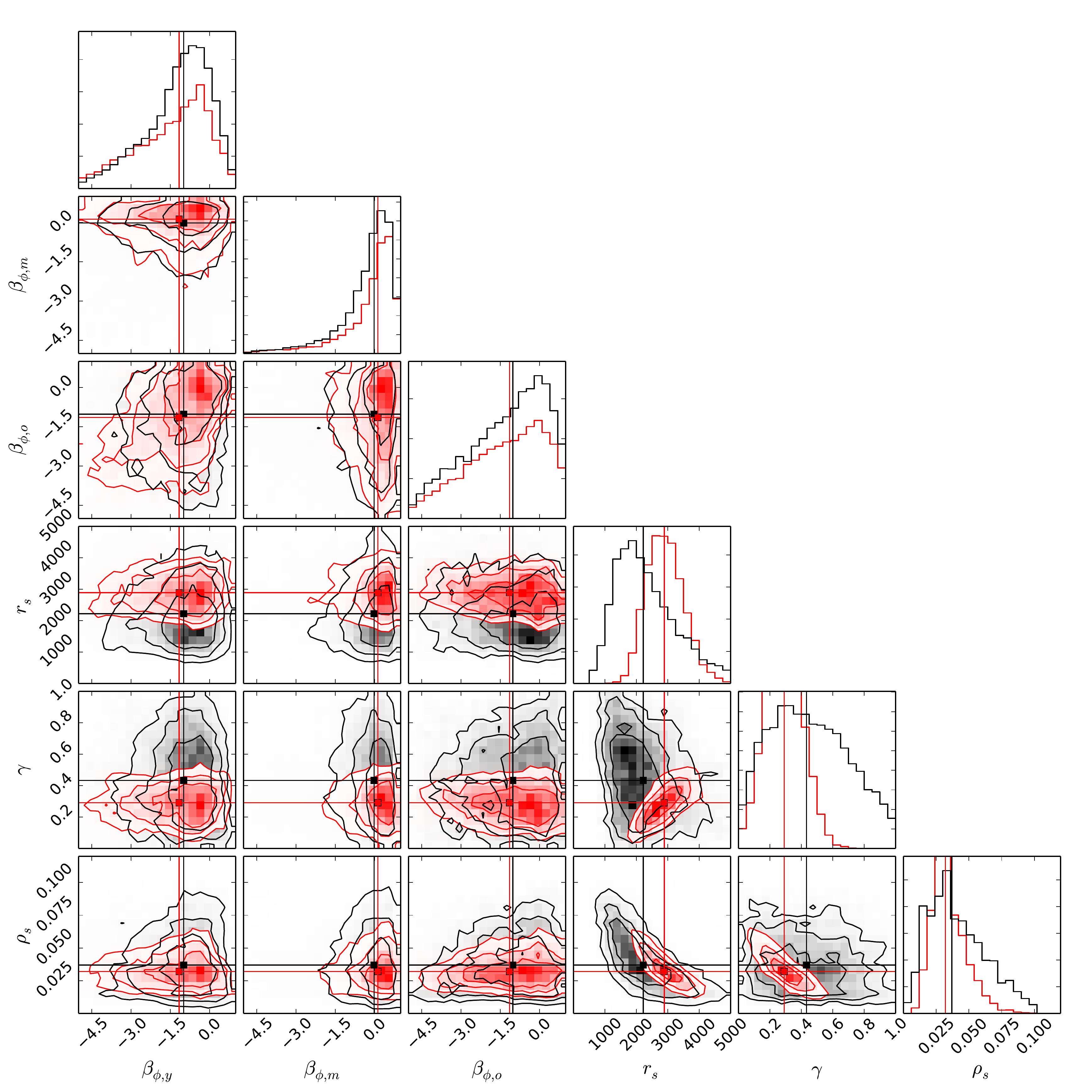}
\caption{Marginalised parameters from the spherical Jeans models: the velocity anisotropy for the young ($\beta_{\phi,y}$), middle-aged ($\beta_{\phi,m}$) and old population ($\beta_{\phi,o}$), and the dark matter parameters$r_\mathrm{s}$, $\gamma$ and $\rho_\mathrm{s}$. Black contours show the marginalised parameter values with Jeans models performed on stellar kinematics only. Red contours show the ones constrained by using $V_\mathrm{c}$ derived from HI kinematics as a prior.
\label{fig_emcee_sph3}}
\end{center}
\end{figure*}

The constrained parameters are plotted in Figure \ref{fig_emcee_sph3}, again with black showing the `Stars only' case and red the `Stars + Gas' case. Compared to the single-population models, only the middle-aged population shows a better constrained $\beta_{\phi,m}$ of $0.13^{+0.48}_{-1.19}$, while both the young- and old-aged populations show similar $\beta_\phi$ of $\beta_{\phi,y}=-1.16^{+1.06}_{-1.82}$ and $\beta_{\phi,o}=-1.15^{+1.34}_{-1.81}$. The derived inner slope of the DM halo in the `Stars + Gas' case is $0.29\pm0.12$, again reaffirming the cored density profile.

\bsp

\label{lastpage}

\end{document}